\documentclass[onecolumn,authoryear]{els-mrw} 

\usepackage{amsmath,amssymb,amsfonts,amsthm,makeidx,graphicx}
\usepackage{txfonts}
\usepackage{subcaption}
\usepackage[inline]{enumitem}
\usepackage{helvet}
\usepackage{hyperref}  
\usepackage{natbib}
\usepackage{wasysym}

\bibpunct{(}{)}{;}{a}{,}{,}

\begin{document}


\chapter{Gravitational wave experiments: achievements and plans}\label{chap1}

\author[1,2]{Elisa Bigongiari}%
\author[1,2]{Matteo Di Giovanni}%
\author[1,2]{Giovanni Losurdo}%

\address[1]{\orgname{Scuola Normale Superiore}, \orgaddress{Piazza dei Cavalieri, 7, 56126, Pisa (Italy)}}
\address[2]{\orgname{INFN}, \orgdiv{Sezione di Pisa}, \orgaddress{Largo B. Pontecorvo 3, 56127, Pisa (Italy)}}

\articletag{Chapter Article tagline: update of previous edition, reprint.}

\maketitle

\begin{abstract}[Abstract]
	Gravitational wave (GW) experiments have transformed our understanding of the Universe by enabling direct observations of compact object mergers and other astrophysical phenomena. This chapter reviews the concepts of GW detectors, such as LIGO, Virgo, and KAGRA, and describes their operating principles, data acquisition and analysis techniques, and some of the methods used to extract source properties. The scientific impact of GW observations is discussed as well, including contributions to astrophysics, tests of general relativity, and cosmology. We also examine the role of multimessenger astronomy and the complementarity between different GW detectors and with other astroparticle experiments. Finally, we outline future prospects with next-generation detectors, like the Einstein Telescope and Cosmic Explorer, and space-based missions.
\end{abstract}

\begin{keywords}

 	Gravitational waves\sep experiments\sep detectors \sep data analysis \sep astrophysical sources
\end{keywords}

\begin{figure*}
     \centering
     \begin{subfigure}[b]{0.4\columnwidth}
         \centering
         \includegraphics[width=0.9\textwidth]{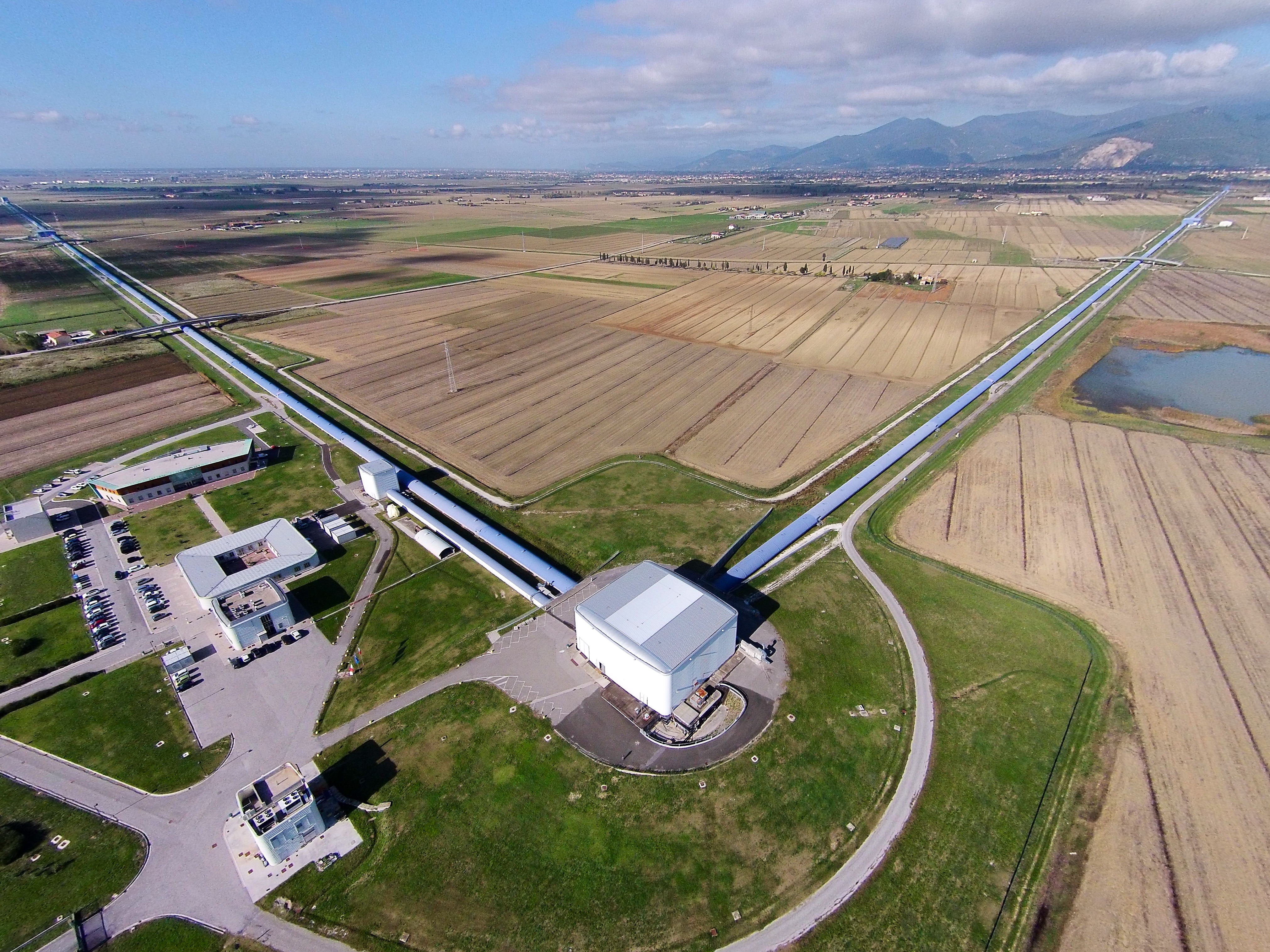}
         \caption{Aerial view of the Virgo interferometer located in Cascina, near Pisa, in Italy (picture courtesy of the Virgo collaboration).}
         \label{fig:snra2}
     \end{subfigure}
     \hspace{20pt}
     \begin{subfigure}[b]{0.4\columnwidth}
         \centering
         \includegraphics[width=0.9\textwidth]{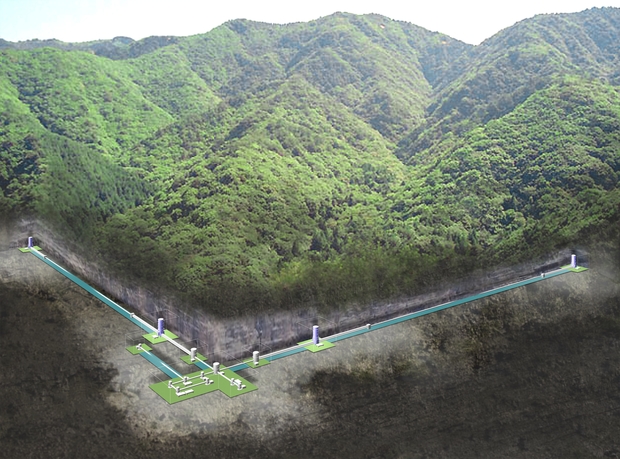}
         \caption{Illustration of the underground KAGRA detector, located in the former Kamioka mine in the Toyama prefecture in Japan (picture courtesy of the ICRR, Univ. of Tokyo).}
         \label{fig:snrb2}
     \end{subfigure}
    \vskip\baselineskip
     \begin{subfigure}[b]{0.4\columnwidth}
         \centering
         \includegraphics[width=0.9\textwidth]{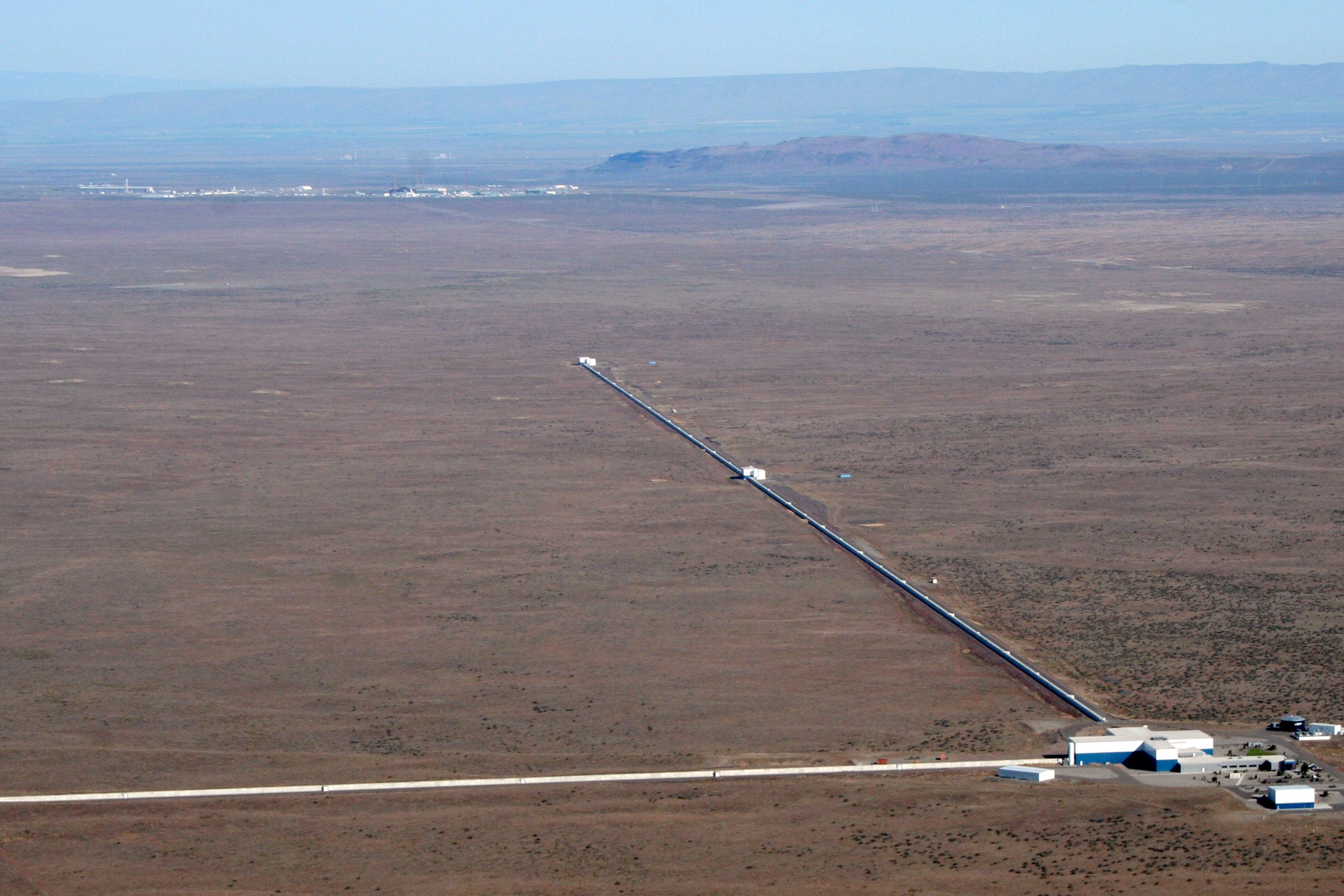}
         \caption{Aerial view of the LIGO Hanford detector, located in Washington State in the USA (picture courtesy of the LIGO Collaboration).}
         \label{fig:snrc2}
     \end{subfigure}
    \hspace{20pt}
     \begin{subfigure}[b]{0.4\columnwidth}
         \centering
         \includegraphics[width=0.9\textwidth]{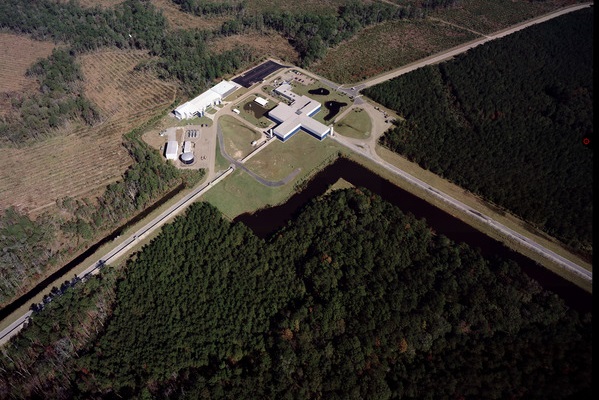}
         \caption{Aerial view of the LIGO Livingston detector located near Baton Rouge in Louisiana in the USA (picture courtesy of the LIGO Collaboration).}
         \label{fig:snrd2}
     \end{subfigure}
        \caption{The four currently active detectors in the current gravitational wave detector network.}
        \label{fig:sites}
\end{figure*}

\begin{glossary}[Nomenclature]
	\begin{tabular}{@{}lp{34pc}@{}}
BH & Black Hole\\
BBH & Binary Black Hole\\
BNS & Binary Neutron Star\\
CBC & Compact Binary Coalescence\\
CDM & Cold Dark Matter\\
CE & Cosmic Explorer\\
CGWB & Cosmological Gravitational Wave Background\\
CMB & Cosmological Microwave Background\\
DE & Dark Energy\\
DM & Dark Matter\\
EM & Electromagnetic\\
EMRI & Extreme-Mass-Ratio Inspirals\\
EoS & Equation of State\\
ET & Einstein Telescope\\
GR & General Relativity\\
GRB & Gamma Ray Burst\\
GW & Gravitational Wave\\
GWOSC & Gravitational Wave Open Science Center\\
IMBH & Intermediate Mass Black Hole\\
KAGRA & Kamioka Gravitational Wave Detector\\
LIGO & Laser Interferometric Gravitational Observatory\\
LVK & LIGO-Virgo-KAGRA\\
NS & Neutron Star\\
PBH & Primordial Black Hole\\
PSD & Power Spectral Density\\
PTA & Pulsar Timing Array\\
SMBH & Supermassive Black Hole\\
SN & Supernova\\
TOV & Tolman-Oppenheimer-Volkoff\\
ULDM & Ultra Light Dark Matter\\
	\end{tabular}
\end{glossary}

\section*{Objectives}
\begin{itemize}
	\item Summarize the history of GW detection 
	\item Outline the working principles of current GW detectors
	\item Give an overview of how to extract the information about the physics of the sources for some specific GW events
    \item Discuss the possible synergies with other experiments in the field
    \item Present the future development in GW experiments
\end{itemize}


\section{History of the experiment}\label{history}

Gravitational waves (GWs) are wavelike perturbations that propagate through spacetime at the speed of light. They are induced by asymmetrically accelerating systems and can be pictured as ripples in spacetime, traveling from the source to the observer while generating alternating compressions and expansions in orthogonal directions. They were first theorized in 1916 by Einstein, who found that the field equations of General Relativity (GR, \cite{feld, GeneralRelativity}) had wave solutions \citep{gravitationalwaves}. Einstein also predicted that the amplitude of these waves was extremely small, making their detection extremely difficult. As a consequence, for one hundred years, the existence of GWs had only been theorized and indirectly confirmed, as in the case of the binary system PSR B1913+16 \citep{PSRB1913+16}. 

\noindent For these reasons, September 14th, 2015, marked a milestone in the way we observe the Universe as GWs were detected for the first time, confirming the theoretical predictions and models initially proposed. On that day, the Advanced LIGO detectors \citep{aLIGO}, located in the USA, observed the GW emission from the merger of a binary black hole system (BBH) with a total mass of $\simeq 65 M_{\bigodot}$ \citep{GW150914}: a new observational window onto the dark Universe was opened and the way in which we observe astrophysical phenomena permanently changed. Later, the two LIGO detectors were joined by the Virgo detector \citep{aVirgo}, located in Italy, and the three of them combined contributed to some breakthroughs in GW detections, such as the first 3-detector observation of a BBH merger \citep{GW170814}, the first detection of a binary neutron star (BNS) merger \citep{GW170817}, the first detection of an intermediate-mass BH (IMBH) \citep{GW190521}, and the detection of a possible Neutron Star-Black Hole (NSBH) merger \citep{NSBH}. Currently, the GW detector network is completed by the Kamioka Gravitational Wave Detector (KAGRA) detector \citep{kagra}, located in the Kamioka mine in Japan. 

\begin{figure}[htbp]
	\centering
	\includegraphics[width=0.3\textwidth]{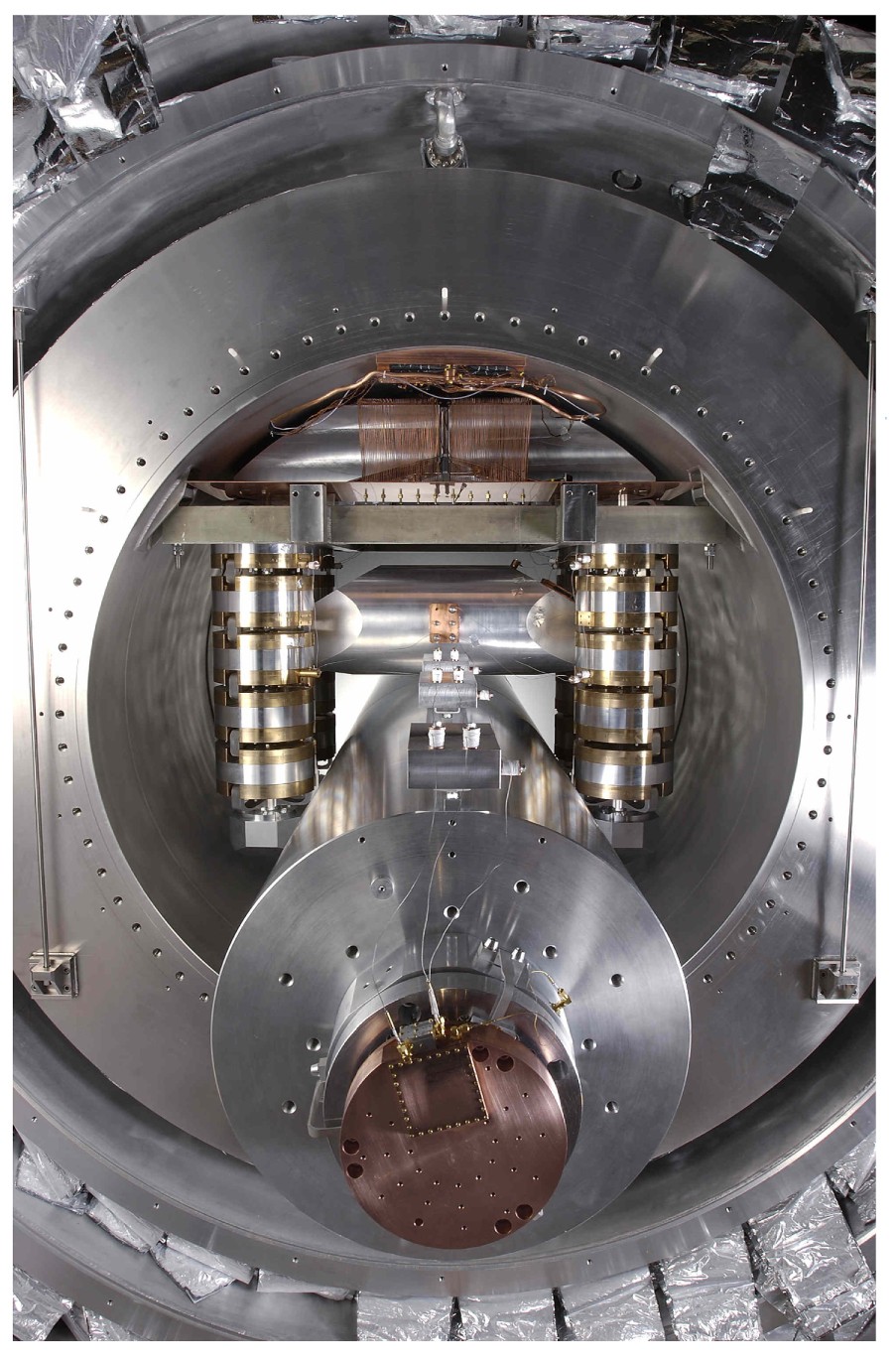}
	\caption{View of the inside of the AURIGA detector, an example of the early GW detectors, located at the INFN Laboratori Nazionali di Legnaro, Italy. The resonant bar antenna itself is the silver cylinder at the forefront of the image. During operations, the antenna was enclosed in a cryostat. Credits: \citep{aurigafig}.}
    \label{fig:AURIGA}
\end{figure}

\noindent Despite the relatively recent discoveries, the history of GW detections can be traced back to the late 1960s, when Prof. William J. Weber designed and built the first prototype of a resonant bar antenna \citep{Weber}. The idea behind this type of detector was to have an aluminium cylinder that would resonate at the passage of GWs. A series of piezoelectric sensors would then convert the vibration signal into an observable electric signal (Figure \ref{fig:AURIGA}). Since then, and until the end of the twentieth Century, with several improvements to the original design by Weber, resonant bar antennas have been the GW detector of choice \citep{nautilus, explorer, auriga, allegro}. When, at the beginning of the twenty-first Century, it became clear that interferometric detectors (Section \ref{generic}) were more promising, resonant antennas were phased out and ceased operations in the 2010s, after supporting the observing runs of the early interferometric detectors.

\noindent The history of GW interferometers can be traced back to the 1960s as well, when the idea of using a Michelson interferometer to detect GWs was first proposed in 1962 \citep{Gertsenshtein:1962}. Nevertheless, a detailed feasibility study, including noise analysis and sensitivity estimates, was carried out only in 1972 by Rainer Weiss \citep{Weiss:1972}. In 1989, after years of conceptual development and R\&D, two major proposals to build large-scale interferometric detectors were submitted:

\begin{itemize}
    \item the Laser Interferometer Gravitational wave Observatory (LIGO) project \citep{LIGO:1989}, in the United States, submitted to the National Science Foundation, proposing the construction of two identical $4 \, \mathrm{km}$-long interferometers separated by $\sim 3000 \,\mathrm{km}$;
    \item the Virgo project \citep{VIRGO:1989}, in Europe, submitted to the French CNRS and the Italian INFN, aiming to build a $3 \, \mathrm{km}$-long interferometer in Italy.
\end{itemize}

\noindent These projects promised to increase the detection bandwidth and sensitivity, and provide access to a wider range of GW sources compared to resonant bar antennas. LIGO and Virgo were formally approved in 1992 and 1994, respectively, and completed in 1999 and 2003. The sites selected to host the LIGO detectors were the plains of the Hanford site (Washington State, USA), a former nuclear production complex of the Manhattan Project, and Livingston (Louisiana, USA), a few kilometers away from the capital of Louisiana, whereas Virgo was placed near Pisa, in Italy. During the first years of operations, often referred to as the initial detector era, LIGO and Virgo were also complemented by two smaller interferometers: GEO600 \citep{GEO}, located near Hannover (Germany), and TAMA300 \citep{TAMA}, located near Tokyo (Japan). During the first decade of the twenty-first century, LIGO and Virgo progressively improved their sensitivity, ultimately reaching their initial design goals and achieving stable operation. Although no detections were made with the initial configurations, these instruments successfully demonstrated the viability of the interferometric technique and laid the technological groundwork for future observations.

\noindent A crucial step forward occurred in 2007, when LIGO and Virgo, initially conceived as separate and, to some extent, competing experiments, signed a landmark agreement for full data sharing and joint publication of scientific results. This collaboration marked the beginning of a true global network of interferometric detectors, designed to operate in a fully coordinated manner, effectively functioning as a \emph{single machine} \citep{MoU:2007}. The synchronization of observational runs and the combination of data streams from the two experiments eventually led to a significant enhancement in the network's ability to localize GW sources in the sky and increase its detection confidence. This shift from competition to cooperation proved essential in establishing GW astronomy as a mature and global scientific enterprise.

\begin{figure}[htbp]
	\centering
	\includegraphics[width=\textwidth]{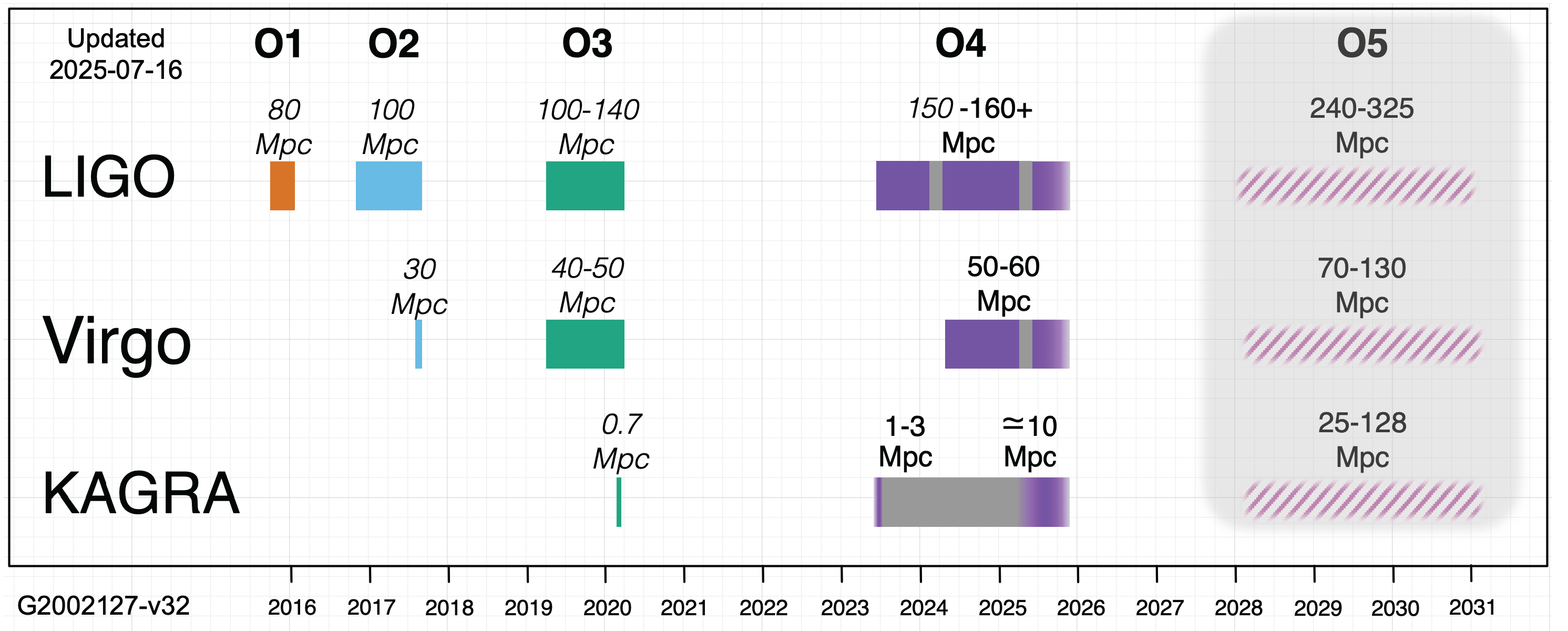}
	\caption{Timeline of the LIGO-Virgo-KAGRA (LVK) operations. Vertical grey bands depict downtime for upgrades and commissioning. The BNS range, i.e., the maximum distance over which each detector can detect a BNS merger with a fixed intensity threshold, is also reported, in different colors.
    Image from the $\,$\href{https://observing.docs.ligo.org/plan/}{IGWN | Public Alerts User Guide}.}
    \label{fig:ObservingRuns}
\end{figure}

\begin{figure}[t]
	\centering
	\includegraphics[width=0.6\textwidth]{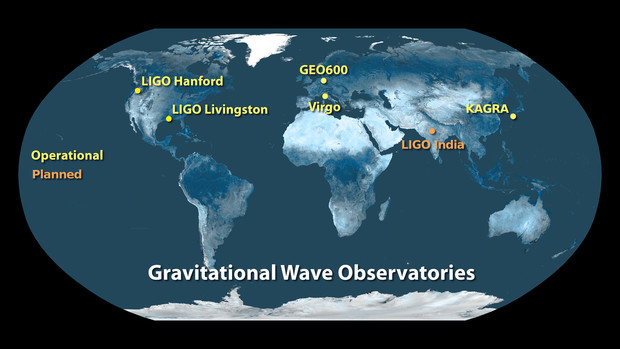}
	\caption{Map of GW observatories across the globe: in yellow are depicted the current (as of 2025) operating facilities, in orange, the planned project of the third LIGO observatory, to be sited in India (Section \ref{future:ground}).
    Image from: \href{https://www.ligo.caltech.edu/}{LIGO Lab | Caltech | MIT}.}
    \label{fig:GWobservatories}
\end{figure}

\noindent The initial era of GW detectors officially ended in 2011, when Virgo was shut down (LIGO had ceased operations a year earlier) to start working on the Advanced generation of detectors. In fact, although initial detectors proved a technical success, they were limited in sensitivity, with the consequence of not delivering the much sought-after first GW detection. Namely, even if the sensitivity of initial LIGO and Virgo would have still been high enough to detect GWs from a supernova (SN, see Section \ref{data:sources}), no such event happened during their years of operations; therefore, an increase of the detectors sensitivity was needed and a series of improvements in the suspensions, optics, and control systems of the interferometers were performed, leading to the Advanced generation era. This era officially began in September 2015 with the operation of the two LIGO detectors and was immediately crowned with success. Virgo would join two years later, while KAGRA became operational in 2020 and joined part of the third and fourth observing runs of LIGO and Virgo. Unlike LIGO and Virgo, the KAGRA detector, with its $3 \, \mathrm{km}$ arms, is built underground and operates at cryogenic temperatures to reduce thermal and Newtonian noise contributions, respectively (Section \ref{generic:noise}). Finally, the already mentioned GEO600, despite being smaller and less sensitive, is still active and is mainly used as a research and development platform for new technologies, but occasionally supports the observations of the LVK network (see Figure \ref{fig:GWobservatories} for a summary). 

\noindent Overall, the current detectors network carried out four observing runs (O1, O2, O3, and O4) of increasing length (Figure \ref{fig:ObservingRuns}): to make a comparison, if the first observing run (O1) lasted only $\sim 3$ months, at completion in November 2025, O4 will have collected a total of almost 2.5 years of data. During this 8-year observation period, almost 250 confirmed GW signals have been detected \citep{gwtc4}, along with nearly 150 significant detection candidates, which will be discussed in future signal catalogs.\\

\section{Generic experiment description}\label{generic}

\begin{figure*}[t]
     \centering
     \begin{subfigure}[b]{0.4\columnwidth}
         \centering
         \includegraphics[width=1\textwidth]{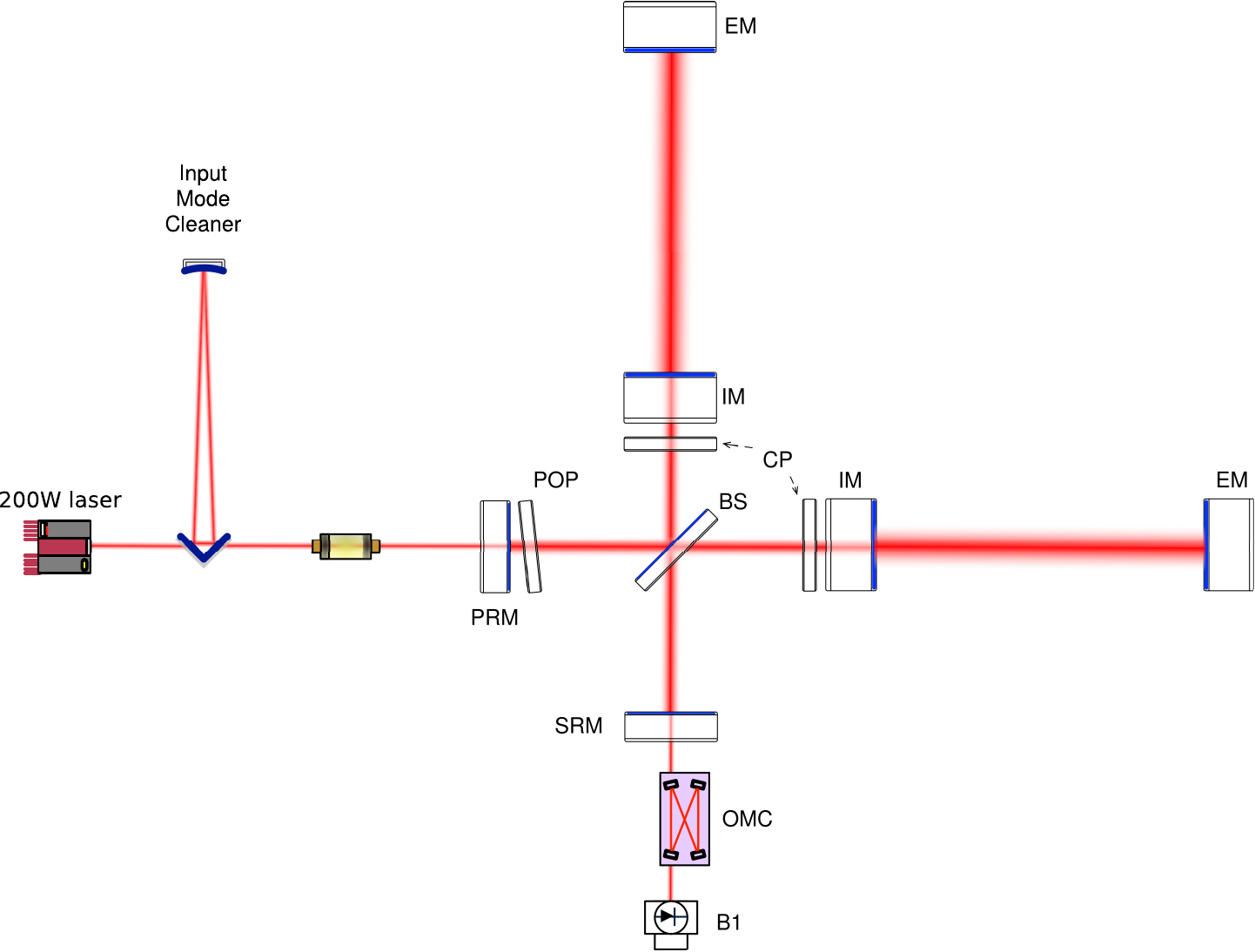}
         \caption{}
         \label{fig:AdV}
     \end{subfigure}
     \hspace{20pt}
     \begin{subfigure}[b]{0.4\columnwidth}
         \centering
         \includegraphics[width=0.9\textwidth]{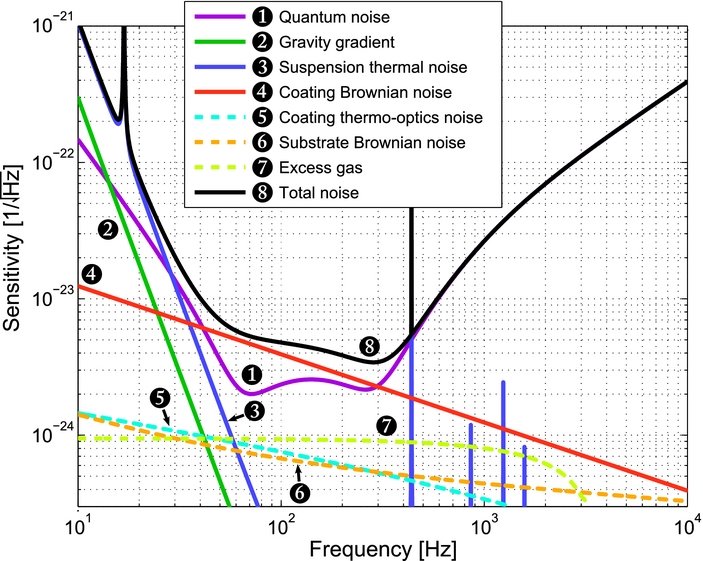}
         \caption{}
         \label{fig:noise}
     \end{subfigure}
        \caption{(a) Simplified optical layout of the Advanced Virgo detector. The laser beam is split by a beam splitter (BS). Each 3 km-long arm cavity is formed by an input mirror (IM) and an end mirror (EM). The power recycling mirror (PRM) and the signal recycling mirror (SRM) are shown along with auxiliary optical components Credits: \citep{aVirgo}. (b) Representative design of the Advanced Virgo sensitivity and noise limitations.
    The reference sensitivity (solid black) is given by the square root of the power spectral density of the detector and consists of the sum of various contributions affecting different frequency ranges (solid and dashed, colored). Credits: \citep{ASD}.}
        \label{fig:sites}
\end{figure*}

As we introduced in Section \ref{history}, the basic operating principle of current GW detectors is that of a Michelson interferometer. In this scenario, the deformations of spacetime induced by the passage of a GW can be detected as length measurements: the distance $L$ between two test masses is subject to fluctuating variations $\Delta L$, which can be measured and associated with the intensity and time evolution of the passing GW. For a typical GW detected by terrestrial interferometers, however, $\Delta L$ is extremely small, up to $\sim 10^{-17} \, \mathrm{m}$, and it clearly cannot be measured directly (as a comparison, the diameter of the nucleus of the Hydrogen atom is $1.68\times 10^{-15}\, \mathrm{m}$). This is why GW detectors use interferometry: by introducing a laser beam propagating between the two test masses (the interferometer's reflective mirrors), one can aim to detect any minor displacement from the phase evolution of the beam.\\
In Figure \ref{fig:AdV}, the Michelson configuration of the Advanced Virgo detector is shown. The laser beam (a single-frequency 1064 nm Nd:YAG laser for LIGO and Virgo) travels along one of the interferometer's arms until it reaches a partially reflective mirror, the beam splitter, which splits the beam into the two perpendicular arms, each several kilometers long. The beams travel down the arms, reflect off the test mass mirrors, and recombine at the output, where a photodetector is placed. In the absence of any GW and neglecting local disturbances, the two beams interfere destructively, so no power reaches the photodetector. The passing of a GW, on the other hand, effectively stretches one arm while compressing the other. This differential deformation changes the optical path lengths, introducing a phase shift between the returning beams, which can be measured as the resulting interference pattern at the photodetector and is directly proportional to the GW-induced strain, $h \sim \Delta L /L$.

\noindent To achieve the sensitivity required to detect such tiny signals, typically $h \sim 10^{-21}$, modern interferometers incorporate several key enhancements to the basic Michelson design. Each arm contains a high-finesse Fabry–Pérot cavity, an optical cavity formed by an input mirror and the end mirror, inside which the beams undergo several consecutive reflections, increasing the effective optical path length and amplifying the GW-induced phase shift by factors of several hundred. A power-recycling mirror at the input port forms a resonant cavity that increases the circulating power, enhancing tens of watts of laser input to hundreds of watts at the beam splitter, and yielding circulating powers of several hundred kilowatts in the arm cavities. At the output port, a signal-recycling mirror forms another resonant cavity with the interferometer arms, allowing the frequency response to be tuned and the detection bandwidth to be broadened.

\subsection{Noise budget of the detector} \label{generic:noise}

The design sensitivity of a GW detector is given by the sum of different noise contributions. Usually, we classify noise sources into fundamental, technical, and environmental noises. Fundamental noises can be computed from first principles, and they determine the ultimate design sensitivity of the instrument (Figure \ref{fig:noise}). It is not possible to reduce this kind of noise without a major instrument upgrade, such as the installation of a new laser or the fabrication of better optical coatings for the mirrors. Technical noises, on the other hand, arise from electronics, control loops, charging noise, and other effects that can be reduced once identified and carefully studied. Environmental noises include seismic motion, acoustic, and magnetic noises. In this Section, we will limit the discussion to fundamental noises for simplicity. A full overview of other noise sources that can affect the operations of a GW interferometer can be found in, e.g., \cite{noises1,noises3,noises2}.

\noindent Noise sources can be divided into classes according to their origins and coupling mechanisms as well. We can split them into displacement and sensing noises: displacement noises are those that cause a real motion of the test masses or their surfaces. On the other hand, sensing noises limit the instrument's ability to measure test mass motion. Nevertheless, this distinction is not perfect, since some noise sources can belong to both categories. In general, the noise sources that contribute to forming the characteristic sensitivity curves of GW interferometers are (refer to Figure \ref{fig:noise}):
  \begin{enumerate}[series=MyList,leftmargin=0ex]
    \item \textbf{Quantum noise} - Also known as \textit{shot noise}, is given by the uncertainty in the number of photons, whose distribution is ruled by a Poisson statistic, that are collected by the photodiode. This contribution can be reduced by increasing the laser power. Nevertheless, an excessive laser power can cause a change in the position of the mirrors due to the momentum transfer from the photons to the optical surface, the \textit{pressure radiation noise}.
    \item \textbf{Gravity gradient} - Also known as \textit{Newtonian Noise}, is not yet limiting the sensitivity of current detectors, but will be relevant for the Einstein Telescope (Section \ref{future}). It is caused by local mass density fluctuations that alter the way in which the test masses interact with the local gravitational field, producing unwanted motions of the mirrors.
  \end{enumerate}
 \begin{enumerate*}[resume=MyList, before=\hspace{-3ex}]
    \item \hspace{0.1ex}
    \item \hspace{0.1ex}
    \item \hspace{0.1ex}
    \item \textbf{Thermal and Brownian noises} - These noise sources originate from mechanical systems (mirrors, suspensions, glass fibers, coating of the mirrors) that can be seen as dissipative oscillators and subject to random thermal fluctuations. The thermal equilibrium with the external environment will make them vibrate, causing an uncertainty in the measurement of their position according to the fluctuation-dissipation theorem \cite{teoF&D}. This noise can be reduced using mechanical systems with low internal dissipations or, as in the case of KAGRA, using mirrors at cryogenic temperatures.
  \end{enumerate*}
 \begin{enumerate}[resume=MyList,leftmargin=0ex]
    \item \textbf{Excess gas} - Refers to the noise generated by residual gas in the vacuum system. This gas interacts with the main laser beam, scattering light into unwanted directions. The scattered photons can then bounce off vibrating surfaces and re-enter the main beam, creating a noise signal that can either mimic a GW or interfere with the detector's control system. For this reason, all critical components are housed in ultra-high-vacuum enclosures to eliminate noise from residual gas motion and acoustic coupling as well.
  \end{enumerate}

\noindent Another noise source that is worth mentioning is seismic noise. This is the main noise source limiting terrestrial GW detectors at low frequencies \citep{digiovanni2025}. Seismic noise, which is induced by any mechanical vibration of the ground (due to weather conditions, earthquakes, infrastructures, etc. \citep{digiovannietal2021, digiovanni2023}), produces variations in the optical path length of the recombined beam, mimicking the passage of a GW. Seismic noise is unavoidable on Earth, and its effects are attenuated by suspending the test masses to multi-stage, high-performance vibration isolation systems, in the so-called inverted pendulum configuration.

\section{From data to physics}\label{data}

The process that allows for obtaining physical observables, such as the source properties or the significance of an event, from the interference pattern formed in an interferometer typically takes place in a matter of minutes. However, it consists of extremely challenging and laborious steps, which we will attempt to summarize in the most comprehensible way in this Section.\\
First of all, constant monitoring of the gravitational data and the status of the interferometers is needed. The real-time monitoring process is set up to respond whenever a gravitational signal occurs in the data in the form of phase shifts in the output beam. Once the trigger happens, the signal is cleaned of any background noise, allowing us to access the source parameters from the extracted signal shape. However, this extraction process must be carried out differently depending on the nature of the GW source, which significantly affects the signal shape. For this reason, different pipelines searching for different signals are always active in data monitoring, complementing each other to cover the plethora of GW sources as widely as possible. These pipelines can also be used after the data have been collected, to eventually refine the source parameters using more computational resources. Sources that require longer observing times have dedicated offline pipelines, which operate only after the data are collected.

\subsection{Zoology of GW sources} \label{data:sources}
As described in Section \ref{history}, any system experiencing an asymmetrical acceleration loses energy in the form of GWs and can therefore be classified as a GW source. Since the time evolution of the asymmetry is reflected in the emitted waveform, different astrophysical systems are expected to produce signals with varying durations, time evolutions, amplitudes, and other characteristics, which need to be investigated using appropriate techniques. This means that the nature of potential GW sources can be varied and their number extremely high. For this reason, we focus here only on the most common sources addressed by current experiments and extend the overview of the science case of GW detectors in Section \ref{physics}. For a more general overview of the principles of GW search methods, however, we invite the reader to look up \citep{AugerPlagnol2017, Maggiore2007, KenathSivaram2023}, and the other papers referenced in this Chapter.

\begin{figure}[t]
	\centering
	\includegraphics[width=0.8\textwidth]{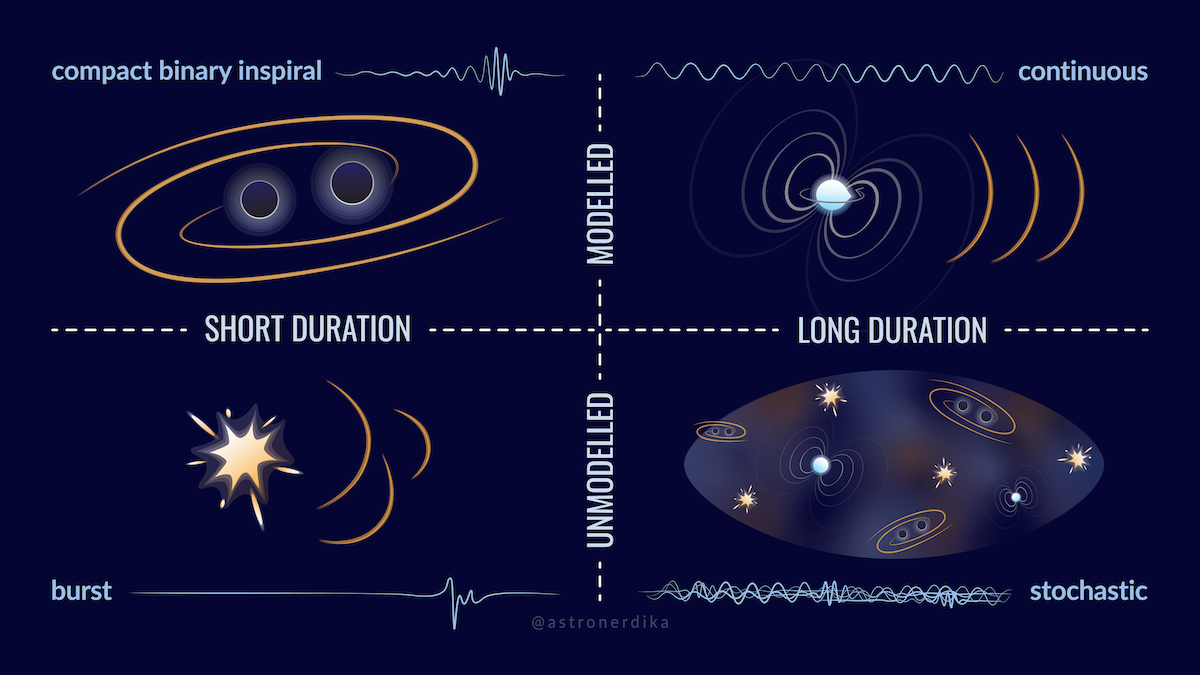}
	\caption{Visual schematization of the four main classes of GW sources and the type of signal produced. On the left, the short-duration signals, divided into CBC (top) and burst (bottom) signals. On the right, long-duration signals, divided into continuous waves (top) and the stochastic background (bottom). The kinds of searches typically performed for the detection of each source class are also shown. Credits: Shanika Galaudage}
	\label{fig:GWsources}
\end{figure}

\noindent A first distinction among GW sources and their expected GW emission can be made in terms of the time duration of the signal emitted (see Figure \ref{fig:GWsources} for a summary):

\begin{itemize}
    \item Transient sources are those expected to undergo asymmetric accelerations inside a limited time window. This can happen for various reasons; typically, when a physical phenomenon affects our system for a limited time span, during which the sought-after asymmetry is induced in the system. Consequently, the signal emitted will have a finite time duration. If the emission happens suddenly and has a very short time duration, the signal can be further classified as a burst signal. This is the case for asymmetric SN explosions \citep{SN} or pulsar glitches \citep{NSglitches}. Generally speaking, the signal evolutions are hard to model due to the extreme width of the space parameters: the waveforms reflect the transient accelerating process that powered the emission, which cannot be uniquely defined. In fact, one can think of several different processes that can induce a burst GW emission, each depending on a varied number of parameters. For each conceivable accelerating process, different values assumed by each of these parameters inevitably produce a different waveform. For this reason, to detect burst signals, the so-called unmodeled searches need to be performed, which are generic searches that do not require any prior assumption on the signal shape and are optimal to search for sources that are hard to model, e.g., SN, unexpected sources, and new physics.\\
    The opposite scenario happens for compact binary coalescences (CBCs): when two compact objects, i.e., astrophysical objects with very high masses confined in a minimal volume, start spiraling toward the common center of mass, they are bound to merge into a single final object. During this whole process, GWs are emitted until the final object reaches a stationary state. By modeling the kinematics of these systems, we can therefore reconstruct the expected waveforms and use them in the signal extraction process, as is done in modeled searches, which we will describe in Section \ref{data:extraction}. In fact, CBCs are the only systems for which GWs have been detected up to now \citep{gwtc4}.

    \item Continuous sources are those systems with continuous asymmetries, which induce the emission of a GW with no specific time duration. This happens, for example, for asymmetrically-rotating NSs: the asymmetry can be either structural, if the star is not perfectly spherical (high ellipticity, local surface deformations, etc.), or kinematic, for instance, if the rotation shows no precession or the star is not revolving in a binary system. In all these cases, the systems are intrinsically asymmetric; therefore, as the star keeps rotating, GWs are emitted continuously with time with almost monochromatic energies and frequencies associated with the star’s rotational frequency \citep{RotatingNS}. According to the information available for the source, these signals can be inquired about either with a modeled or an unmodeled search \citep{PulsarO4}.
    Another example of long-duration GWs is the stochastic GW background, which we will describe in depth in Section \ref{physics:cosmology:universe}. It consists of the superposition of multiple incoherent unresolved GWs of various potential origins, and, for this reason, it can only be queried with unmodeled searches \citep{GWB}.
\end{itemize}

\subsection{Data acquisition} \label{data:acquisition}
When an interferometer is taking data, or as we like to say, is in science mode, any length variation $\Delta L(t)$ in the detector arms, which is converted into phase shifts $\Delta \phi(t)$ in the recombined beam, is recorded digitally by the data acquisition system. As we introduced in Section \ref{generic}, the time evolution of such deformations can be converted into a dimensionless factor, which we will denote here as $d(t)$:

\begin{equation}
    d(t)=\frac{\Delta L(t)}{L} = \frac{\Delta\phi(t)}{\phi}
\end{equation}

\noindent Ideally, the time series of the collected data $d(t)$ would indicate exactly the time evolution of the amplitude strain $h(t)$. In practice, however, noise contributions $n(t)$ of various natures are always included so that:

\begin{equation}
    d(t) = n(t) + h(t)
\end{equation}

\begin{figure}[t]
	\centering
	\includegraphics[width=\textwidth]{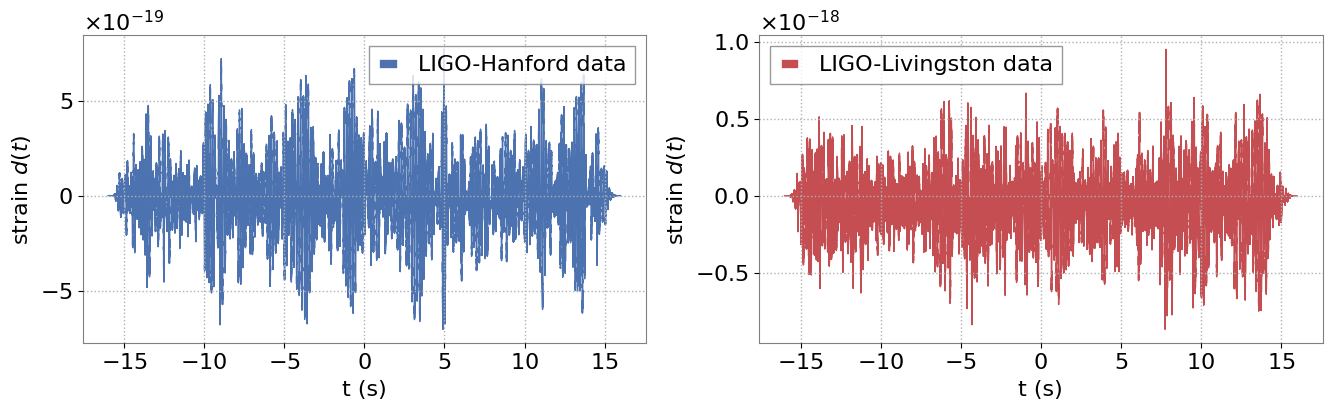}
	\caption{32 seconds of raw data $d(t)$ centered at the trigger time of event GW150914, including both the waveform $h(t)$ and the noise $n(t)$ from the two LIGO interferometers, LIGO-Hanford (blue) and LIGO-Livingston (red). Real data taken from the \href{https://gwosc.org/}{Gravitational Wave Open Science Center (GWOSC)}.}
	\label{fig:GW150914}
\end{figure}

\noindent To access any potential signal $h(t)$, we must therefore characterize and then subtract the noise contributions affecting different frequencies. These contributions to the collected data can be very intense and usually bury any gravitational strain, making it completely impossible to identify the signal by simply looking at the time evolution of the data (see Figure \ref{fig:GW150914} as an example).

\subsection{Noise characterization} \label{data:noise}
As we have already mentioned earlier in the text, not all noise sources of a GW detector are predictable. As a consequence, understanding the noise behavior of the detector during an observing run is crucial for extracting any gravitational signal and consequently accessing the source properties. To do that, it is convenient to work in the frequency domain, instead of working with the time series $n(t)$. We therefore use the Power Spectral Density (PSD, $S_n(f)$), which is defined as the square module of the Fourier transform of the noise: 
\vspace{-3mm}

\begin{equation}
    S_n(f) = \lim_{T\to \infty} \frac{1}{T} \left| \int_{-T/2}^{T/2}n(t) \cdot e^{-2\pi i t} dt \right|^2
\end{equation}

\noindent Namely, this quantifies the variance of the noise fluctuations at a given frequency $f$, around a null mean value. This allows us to understand the impact that noise has on the data inside different frequency ranges: the higher $S_n(f)$ is, the noisier the data taken at the frequency $f$ is. For terrestrial interferometers, the PSD is typically evaluated in the range $f \sim [10-10^3] \, \mathrm{Hz}$, inside which it shows a bucket shape (Figure \ref{fig:noise}). Different frequency ranges are dominated by different noise contributions (Section \ref{generic:noise}): the highest ones are usually found at low frequencies $\sim 10 \, \mathrm{Hz}$ due to seismic effects, where the PSD peaks. The high frequencies $\sim 10^3 \, \mathrm{Hz}$ show quite intense contributions as well, mainly defined by the quantum shot noise. Finally, the PSD is minimal at intermediate frequencies $\sim 10^2 \, \mathrm{Hz}$, where the lowest noise contribution is found, mainly thermal noise.\\
Another approximation we can make is to consider the noise to be Gaussian and assume that the distribution follows a Gaussian distribution, with median 0 and variance $S_n(f)/2$.\\

\subsection{Example of signal extraction in the case of CBC signals} \label{data:extraction}
\noindent Now that we have characterized the noise, we want to compare our data with the expected noise spectrum $S_n(f)$ to extract the potential signal $h(t)$. To do that, we first need to transform the data and move to the frequency domain as well, and we do that, again, with a Fourier transform:
\vspace{-4mm}

\begin{equation}
    d(f) = \int_{-\infty}^{+\infty}d(t) \cdot e^{-2\pi i t} dt
\end{equation}
In general, two main approaches enable the signal extraction: the unmodeled and the modeled searches. 
Looking over all the search methods and pipelines in use in GW searches would require extensive discussion and is beyond the scope of this Chapter. For this reason, in this Section, we focus only on the method used to detect CBC signals. 

\noindent Even if CBC sources are predictable in some sense, the exact gravitational waveform in our data is unknown a priori. For this reason, we make use of signal models, which we will denote as $\bar{h}(t)$, which are artificially constructed waveforms from simulations of various CBC systems. The goal is to find the waveform $\bar{h}_i(t)$ that best models the actual signal $h(t)$ in our data. Once this is done, we will have obtained the real signal and will be able to start estimating the source parameters.\\
\noindent To begin with, a bank of template models is preliminarily built, containing a huge number of waveforms. Each of these is constructed consistently with different source parameters (masses, orbit inclination, arrival time, etc.) so that the entire bank will cover the parameter space as widely as possible.
Now, it may be helpful to imagine that during the whole data-taking period, each of these waveforms is being superposed with the data and, in some way, the level of match between the two is constantly being computed. This process, called matched filtering, allows us not only to trigger a gravitational signal in our data, which happens when a high-level match is reached, but also to access the real strain by considering it as the model that showed the highest match with the data.\\
In practice, the process is more elaborate and requires a statistical approach: the level of match between model and data is quantified by the matched-filter optimal statistic $(d,\bar{h})$, which weights the data with a specific model signal in the frequency domain and normalizes it with the noise spectrum.

\begin{equation}
    (d,\bar{h}) = 4 \int_{0}^{+\infty} \frac{d(f) \hspace{0.5mm} \bar{h}(f)}{S_n(f)} df 
\end{equation}

\noindent Now, if one or more of the optimal statistics we are computing exceeds a fixed threshold, we can expect a gravitational signal to be present in the data. More specifically, if $(d,\bar{h}_i)$ is the highest matched-filter optimal statistic, we can consider $\bar{h}_i$ as the waveform that simulates best the real strain in the data.

\subsubsection{Parameters estimation} \label{data:parameters}
\begin{figure}[h]
	\centering
	\includegraphics[scale=0.25]{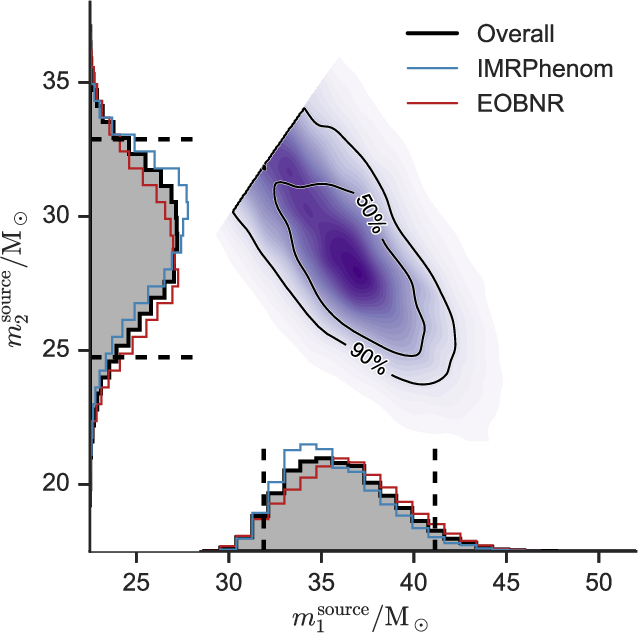}
	\caption{Unidimensional and bidimensional distributions for the two component masses of GW150914. The colored solid lines shape the unidimensional distributions built with two waveform models, IMRPhenom \citep{IMRPhenom} and EOBNR \citep{EOBNR}, while the average of the two is displayed with the solid black line peaking at 
    35.8 $M_{\bigodot}$ and 29.1 $M_{\bigodot}$, respectively. The dashed lines fix the confidence level at 90\%, as described in the text. Credits: \citep{GW150914properties}.} 
    \label{fig:posterior}
\end{figure}

The match-filter optimal statistic is linked to a key quantity, the likelihood ratio $\Lambda(\mathcal{H}_1|\hspace{0.5mm}d)$, which states the likelihood that the data $d(t)$ contains a given signal. Considering that the model $\bar{h}_i$ we have just obtained is shaped by a specific set of still-unknown parameters $\{m_1, m_2, \iota,t_0, ...\}$, we can focus on one of those, for example the inclination of the binary system's orbit $\iota$, and analyze the distribution of $\Lambda$ versus $\iota$. When repeated for each source variable, this process, called marginalization, allows us to determine the best-fit value of each parameter as the value at which the corresponding $\Lambda$ distribution is peaked. This is done by integrating the distribution over every other parameter except the one we are focusing on.
The uncertainty, on the other hand, is fixed accordingly to the confidence level chosen by the width of the same distribution.\\
The accuracy and precision of these source parameter estimates are influenced by the detector's noise, the detector's response to the specific amplitude and phase of the GW, and by the quality of the models used in the search.\\
Usually, multidimensional distributions are considered to take into account potential correlations between different parameters. When considering bidimensional distributions, for example, we focus on how $\Lambda$ changes with two independent parameters $\alpha$ and $\beta$. We do that by integrating over every source parameter except these two: the distribution $\Lambda(\alpha,\beta)$ that we get is therefore a three-dimensional distribution, and it is usually represented as seen from the $\alpha-\beta$ plane. Higher values of $\Lambda$ are generally portrayed accordingly to a specific color scale, so that we can immediately note any correlation between the two parameters (Figure \ref{fig:posterior}).

\subsubsection{The example of GW150914}
Let us now consider the renowned event of the 15th September, 2014 (GW150914, \citep{GW150914}), the very first GW detection ever made.\\
In general, any pipeline that performs modeled searches relies on different sets of waveform models used for the signal matching. \\
Let's assume that we are performing our search with a specific set: first, we will be able to find the model $\bar{h}_i$ that best matches the data. From the model, the corresponding matched filter optimal statistic $(d,\bar{h}_i)$ and the likelihood $\Lambda(\mathcal{H}_1|\hspace{0.5mm}d)$ are built. Finally, we can obtain the likelihood distribution by focusing on one or more chosen parameters; let’s take the first and second component masses $m_1$ and $m_2$, for example. First, we look at the $\Lambda(m_1)$ and $\Lambda(m_2)$ distributions and search for the peak values $\bar{m}_1$ and $\bar{m}_2$. These will be the most likely values for the two parameters. Then, we fix a specific confidence level, let's call this $n$. To assess the uncertainty $\Delta m_{1,2}$ in the parameter estimations with that specific confidence level, we consider the corresponding width of the distribution.\\
For example, for $m_1$:

\begin{equation}
    n = \int_{\bar{m_1}-\Delta m_1}^{\bar{m_1}+\Delta m_1} \Lambda(m_1) \hspace{0.5mm} dm_1 
\end{equation}

\noindent So that $m_1 = \bar{m}_1 \, \pm \, \Delta m_1$. Usually, the confidence levels $n$ are displayed in percentiles, the most frequently used being 90\% or 50\%.

\section{Physics topics addressed by the experiment} \label{physics}
Since the very first detection \citep{GW150914}, GW observations have proven to carry crucial information on several different aspects of physics: from enabling a deeper understanding of astrophysical sources and providing tests of GR \citep{GeneralRelativity} to implications for cosmology. GW detections have therefore paved the way for unprecedented knowledge of the Universe and its evolution, marking the dawn of a new era of scientific exploration. 

\subsection{Astrophysics} \label{physics:astro}

\begin{figure}[t]
	\centering
	\includegraphics[width=0.7\textwidth]{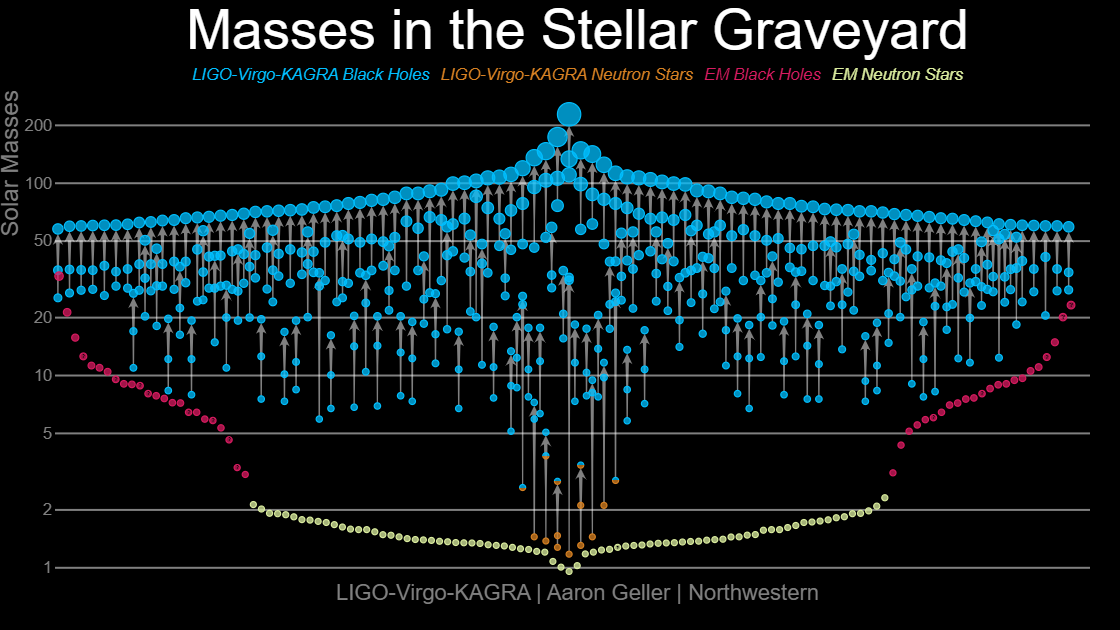}
	\caption{This plot summarizes the masses of BHs and NSs detected with EM observations (magenta and yellow, respectively - updated on January 2024) and with GW (blue and orange, respectively - updated on September 2025). The arrows connect the two merging objects, from the less to the more massive, with the remnant. The vertical axis indicates the mass, in units of solar masses, while the horizontal axis represents no specific quantity. Credits: \href{https://www.ligo.caltech.edu/news/ligo20250826}{LIGO-Virgo | Aaron Geller | Northwestern University}.}
	\label{fig:Graveyard}
\end{figure}

Before the first GW detection, compact objects were only empirically accessible through the detection of their electromagnetic (EM) emissions (e.g., the emission from accretion discs around BHs or NSs in a stellar binary or the radio emission from pulsars, see, for example, \citep{TidalDisruption, PRT}). Therefore, their properties, such as formation rates, host galaxies, mass limits, etc., were the results of theoretical models based only on a set of limited observations. Since GWs can, in general, be emitted and observed from systems without requiring any EM emission, their detection opened the door to previously inaccessible astrophysical objects (Figure \ref{fig:Graveyard}), such as the heaviest stellar BHs, IMBHs \citep{GWTC3}, and, with future ground and space-based detectors \citep{3G}, even supermassive black holes (SMBHs). Old unanswered questions about these objects, such as their origin, mass distribution, interactions, and evolution, started to be investigated in much more detail \citep{BHs} alongside the rise of new scenarios. The morphology and physical properties of BHs and NSs, as remnants of the initial system \citep{NSformation}, were also probed for the first time by the analysis of the post-merging phase. It was possible to shed more light on the structure of NSs and their composition too \citep{Ultra-dense}, by constraining their still unknown equation of state (EoS) \citep{GWs&NSs}. Detections of compact object mergers are also an important tool to probe stellar evolution models and the dynamics of binary formation.
The combination of the multiple GW detections obtained up to now has also allowed us to build catalogs containing the properties of these systems, which can be used, for example, to understand the astrophysical processes (mass transfers, tides, etc.) involved in a coalescence and their impact, by comparing them to predictions of binary evolution models. Estimates of the occurrence rates can also be obtained from these catalogs, and they can be further constrained and refined by assessing more and more detections of this kind.

\subsection{Nuclear physics} \label{physics:nuclear}
The detection of GWs from NSs, whether they are isolated or components of a binary system, can also provide a unique tool to probe the EoS of matter in extreme conditions. In fact, we estimate that the density of a NS is $\sim 10^{15} \,\mathrm{g/cm^3}$, way beyond what can be achieved with nuclear physics experiments on Earth. This means that the behavior of matter at these densities is still unknown. 

\noindent Generally speaking, NSs are thought to be the remnants of the core of a massive star, which first collapsed and then ejected its outer layers through a SN explosion. For this reason, NSs are currently the best-known site for the formation of ultra-dense matter, i.e., a state of matter which appears as compactness increases: quantum effects start emerging, and an equilibrium state is reached as the electron degeneracy stabilizes the structure \citep{Ultra-dense}. As ultra-dense matter cannot be described as a classic fluid like ordinary matter, a specific EoS must be defined from the Einstein equations, by applying some boundary conditions which strongly rely on the model assumed for the description of the NS. Different boundaries, therefore, lead to different EoS \citep{EoS}, all equally valid, having in common a different maximum mass expected for the NS (the Tolman-Oppenheimer-Volkoff, TOV limit), over which the system is inevitably bound to collapse into a BH \citep{T, OV}. We can therefore define a mass gap (the lower mass gap) comprising all the suggested TOV limits inside which a compact object cannot be univocally identified as a NS or a BH. To probe this gap, one can exploit NS observations, specifically mass measurements, to constrain the mass gap more and more. 

\begin{figure}[t]
	\centering
	\includegraphics[width=0.5\textwidth]{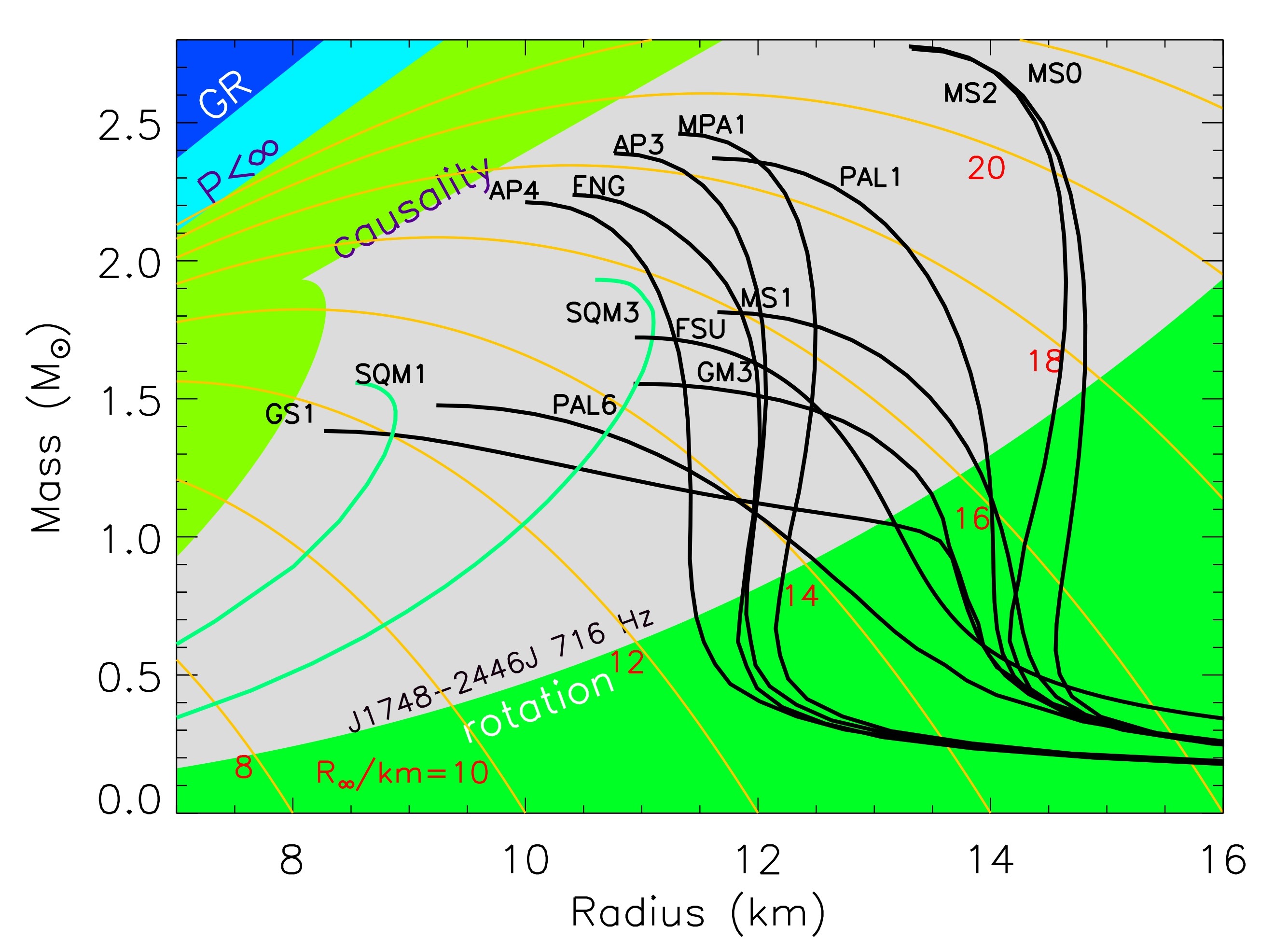}
	\caption{A mass-radius diagram showing different families of NS EoS (black and green lines). Points located above any of these curves are described as BHs by the corresponding EoS, or as NSs otherwise. For more details, see \citep{MassGap}.}
	\label{fig:EoS}
\end{figure}

\noindent Although NSs and their masses could already be accessed with EM observations (pulsar radio timing \citep{PRT} and X-ray spectroscopy \citep{NICER}), GWs provide a complementary and easier way to access and constrain properties of NSs. In fact, the internal structure of a NS can influence its shape and the properties of the crust. This can lead to asymmetries in its structure that generate continuous GWs from rotating isolated NSs. The composition of the interior can also have effects on the waveform of BNS mergers. In fact, unlike BHs, NSs are solid bodies that will deform up to the rupture point during the coalescence, leaving characteristic signatures related to the EoS. Moreover, through waveform reconstruction, we can measure the masses of a BNS before merger, providing even another possibility to test the validity of different EoS (Figure \ref{fig:EoS}) \citep{GWs&NSs}) each of them having different limits on the allowed masses for the NS. Consequently, any EoS that foresees masses lying below the observed values needs to be ruled out. 

\noindent Nevertheless, the full potential of GWs in this field is far from being exploited. The limited number of BNS mergers detected so far \citep{gwtc4} and the lack of detections of continuous GWs from isolated pulsars mean that only loose constraints can be set on the EoS of matter at extreme densities. As we achieve higher sensitivities with current and future detectors, however, we expect to be able to put more stringent and effective constraints on ultra-dense matter properties and on the internal structure of NSs.\\

\subsection{Tests of general relativity} \label{physics:GR}

The specific assumptions on which the principles of GR are based make it possible to put forward alternative theories of gravity to GR, where one or more of these hypotheses fall \citep{GravityFields}. These scenarios were tested long before the discovery of GWs, with several independent experiments that eventually agreed with GR to within percentiles of $\sim 10^{-3}$. These tests mainly consisted of measuring the effects of gravity on specific systems and comparing the results with the theoretical expectations from GR \citep{classicgrtest,classicgrtest2,classicgrtest3,classicgrtest4}. Some of these experiments involved measuring deviations in the propagation of EM radiation or the motion of planets, where the measured deviations are always very small \citep{GRweaktests}. However, all these tests had only been able to probe the theory in the weak-field regime. Since we expect the biggest deviations from GR to manifest themselves when strong gravitational fields are involved, the ideal laboratories for the strong-field regimes are massive, compact objects, such as BHs.
It was only with the first GW detection that we were able to probe this new regime \citep{GW150914Test}. 
To begin with, CBCs are events that cannot be predicted using Newtonian gravity. Only by assuming that part of the system's energy is radiated away as GWs can we expect the orbits to shrink gradually and the merging of the two objects to happen. A Newtonian approach, on the other hand, would not account for any energy emission from the system, which would find itself in a stationary condition where the distance between the two objects remains constant.
Therefore, the detection of a GW confirms, in the first place, the need for a beyond-Newtonian theory, which was long ago formulated to be GR \citep{GeneralRelativity}. What we want to do now is to confirm whether GR, in its own formulation, is an accurate theory for describing these regimes as well, or if some corrections may be needed. To do that, we can perform different tests on GW detections \citep{GRtests}, to name a few:\\

\begin{itemize}
    \item \textbf{Consistency tests:} in this case, the idea is to test the agreement between the observed GW signal and the theoretical models from GR for the expected GW emission. This can be done, for example, by looking at the residual distribution built after subtracting the model $\bar{h}_i(t)$ that best describes the GW strain from the collected data $d(t)$: if GR can describe the evolution of that system, the residual distribution should be consistent with the expected detector's noise distribution $n(t)$ (Section \ref{data:noise}) \citep{BAYESWAVE}.\\

    \item \textbf{Parametrized tests:} to assess the need for any correction in the formulation of GR, one can introduce specific correction parameters in the waveform models. Again, we compare the observed data during the inspiral, merging, and post-merging phases with the corrected models, but this time we want to constrain the correction parameters. If no deviation from GR is needed, each correction should be strictly constrained around zero \citep{ParametrizedTest1, ParametrizedTest2, ParametrizedTest3}.\\

    \item  \textbf{Tests on the remnant properties:} we can also use GR to predict how the excitation modes of the system's remnant will be relaxed during the post-merger (ringdown) phase of a BBH merger. In fact, the nature of the system is expected to influence the nature of the remnant itself \citep{Remnant}. If GR is valid, and the nature of BHs follows Kerr's predicitons \cite{Kerr}, the remnant produces a specific ringdown spectrum. Using post-merger data to infer the final BH area in conjunction with the initial pre-merger area, assessed from the merger signal cycles, makes it possible to test Hawking's BH area law. Recently, both the Kerr nature of BHs and Hawking's area law have been tested with the detection of GW250114, which allowed us to carry on some precise tests of GR \citep{hawking}.\\
\end{itemize}

\noindent All these tests can, in principle, be performed for each GW candidate that is observed. Consequently, by combining results from an increasing number of different detections thanks to upcoming observing runs and new-generation detectors, we expect to obtain stricter tests, higher detector sensitivities, and the development of more accurate waveform models \citep{TestsGWTC3, TestsFuture}.

\subsection{Cosmology} \label{physics:cosmology}


In the previous sections, we have focused on the impact that GWs have on our knowledge of astrophysical sources and how they can be used to probe the validity of GR. In the following, we will extend our description to a new class of GW sources, those with a cosmological origin, i.e., the ones that were emitted long before the formation of stars and galaxies, and therefore encode information on the earliest stages of the Universe. We will emphasize the impact that they have on cosmology in terms of the evolution of the Universe, fundamental physics, and even dark matter.

\subsubsection{Hubble constant} \label{physics:cosmology:H}
In the first decades of the twentieth century, Edwin Hubble \citep{Hubble} and Abbé G. Lemaître \citep{Lemaitre} observed that distant galaxies were moving with a recession velocity $v$, and therefore a redshift $z$, proportional to their distance from the observer $d$:

\begin{equation}
    v = H_0 d \longrightarrow d = \frac{cz}{H_0}
    \label{eq:Hubble_law}
\end{equation}

\noindent This was found to be the consequence of a Universe expanding at a current rate $H_0$, usually referred to as the Hubble constant, whose value is still under investigation. The two main methodologies used over the years have led to inconsistent $H_0$ values. On one hand, observations of the Cosmological Microwave Background (CMB) have yielded a $H_0 = (67.4 \pm 0.5) \, \mathrm{km} \, \mathrm{s}^{-1} \, \mathrm{Mpc}^{-1}$, obtained from the spatial anisotropies of the temperature in the CMB \citep{H0CMB}. On the other hand, measurements of distances and recession velocities/redshift of standard candles, i.e., astrophysical objects with a known intrinsic brightness, resulted in a higher $H_0 = (73.04 \pm 1.04)  \, \mathrm{km} \, \mathrm{s}^{-1} \, \mathrm{Mpc}^{-1}$ \citep{SH0ES}. Although more recent observations have corroborated the former \citep{H02025}, the issue is still open and of major interest.\\
In this scenario, GWs from CBCs can provide us with another independent measurement of $H_0$ and therefore shed more light on the topic. The idea is to combine in Equation \ref{eq:Hubble_law} independent measurements of distances and redshifts for different GW events to obtain $H_0$. However, while parameter estimations yield the value of the source distance, GW detections alone cannot disentangle the redshift from other source parameters. To infer the redshift, and therefore the Hubble constant, we need some independent aids that can either come from the direct observation of an EM counterpart (bright sirens methods), as was the case of GW170817 \citep{H0170817} (Section \ref{compl:mma:170817}), from statistical inference on the mass distribution for CBCs \citep{SpectralSirens}, or by exploiting galaxy catalogs \citep{FirstGalaxyCatalogs}. The last two methods (dark sirens methods) have been performed on data from the third Gravitational-Wave Transient Catalog (GWTC-3), both yielding values of $H_0 \sim 68 \, \mathrm{km} \, \mathrm{s}^{-1} \, \mathrm{Mpc}^{-1}$ \citep{DarkSirens}.

\subsubsection{The early Universe} \label{physics:cosmology:universe}
To probe the early Universe, as described at the beginning of this Section, GWs of cosmological origin must be taken into account. Regardless of their specific origin, they are thought to form what we already introduced as the stochastic GW background, i.e., the combination of multiple incoherent unresolved GWs. It is the detection and characterization of this background that could inform us on the physics associated with the evolution of the early Universe.
This is because EM radiation can only bring us back to $\sim 380^{\cdot}000$ years after the Big Bang, when the Universe stopped being opaque to light and for the first time photons started to propagate freely without scattering with other particles. 
We are therefore unable to access any information regarding earlier ages of the Universe with the detection of photons only. Nevertheless, several models predict the emission of GWs before this era, primarily inflationary models \citep{Inflation}, phase transitions during the Universe's expansion \citep{PhaseTransitions}, and cosmic strings \citep{Strings}. All these sources are expected to produce different GW spectra; therefore, the detection of a gravitational background would allow us not only to observe mere fractions of a second after the Big Bang, but also to understand its origin.

\subsection{Physics beyond the Standard Model} \label{physics:DM}

\subsubsection{Dark Matter}
For many decades, several experimental evidences have established that matter in the Universe is dominated by a stable component, which does not emit light, is non-baryonic, does not carry electric charge, and interacts only through gravity with ordinary matter, i.e., baryonic matter \citep{dm1,dm2}. For these reasons, it is referred to as Dark Matter (DM). Currently, several candidates explain the origin of DM, but none have been detected or ruled out yet \citep{dm1,dm2}. As a consequence, DM is still one of the biggest mysteries in our understanding of the Universe. 
Today, GW searches can potentially contribute to grasping the nature and origin of DM, too, by direct observation of the expected GW signal from some DM candidates. In fact, DM can reveal itself in GW signals in at least two ways, either from direct GW emission from processes involving DM candidates or in traces left in GW signals from other sources \citep{dm6}. Here, we will summarize the most common DM searches in GW experiments.

\noindent A popular opinion, even though not widely accepted, is that a portion of DM is made up of BHs that formed due to density fluctuations caused by an early, rapid expansion of the primordial Universe, called inflation \citep{pbh5,pbh6,pbh1}. Some of these regions were so dense that they collapsed under their own gravity, forming primordial black holes (PBHs). PBHs fulfill all of the requirements to be good DM candidates: they are cold, stable, and non-baryonic, because they formed before the nucleosynthesis \citep{pbh5,pbh6}. Moreover, since their formation is caused by inflation, they can only be explained with physics beyond the Standard Model. In addition to that, if a significant fraction of DM is in the form of PBHs, a portion of them would be in binaries, the coalescence of which would be detectable by ground-based GW detectors \citep{pbh4}.

\noindent Due to their origin, PBHs can have any mass, spanning from a fraction of a Solar mass up to several hundred solar masses. For example, the James Webb Telescope recently announced the discovery of a candidate for a PBH of 50 million solar masses \citep{jameswebb}. As a consequence, a BH could be a candidate for a PBH if it is clear that its mass is outside the typical range of a BH formed by stellar collapse ($M_{BH} \sim [10, \, 50] \, M_{\bigodot}$). Moreover, if multi-Solar-mass PBHs made up the entirety of DM in the Universe, there would be several orders of magnitude more BBH merger events than those observed by LIGO and Virgo \citep{pbh1}. Hence, we can already put constraints on the portion of PBHs in DM using current observations. For example, in the BH mass range accessible to current detectors ($10 \, M_{\bigodot} \lesssim M_{BH} \lesssim  100 \, M_{\bigodot}$), this portion is in the range $[10^{-2},10^{-3}]$ \citep{pbh1,pbh2,pbh3}.

\noindent Another field in which GW searches can contribute to uncovering the nature of DM is the case of bosonic DM. In fact, one of the most common models available to explain the nature of DM involves the existence of ultralight bosons called axions \citep{axions2,axions2,axions3,axions4,axions5,axions6}. If such particles exist, they could appear around rotating BHs due to quantum fluctuations. They would then scatter off, extract angular momentum from these BH, and form macroscopic clouds through a superradiance process \citep{axions7,axions8}. After the BH spin decreases below a fixed threshold, the superradiance process stops and the cloud depletes over time, generating almost-monochromatic, long-duration GWs. 
For this reason, GW detectors could potentially detect evidence of the existence of axions. Boson clouds could also form around BHs in binaries, affecting their spin and the waveform of the coalescence signal. Nevertheless, individual spins are hard to measure, and we have to combine spin measurements from various BBH mergers to obtain constraints on boson cloud/spin interactions \citep{axions10,axions11}. So far, the GW signature of boson clouds around isolated BHs or in BBH mergers has not been detected yet. Nevertheless, it has been possible to set upper limits at a 95\% confidence level on the amplitude of the signal and mass of the axions \citep{axions9, axions10,axions11,axions12, bosonO4}. 

\noindent Finally, bosonic DM can manifest itself in ground-based GW detectors as ultralight dark matter (ULDM) as well. ULDM is a class of DM models, where DM is composed of bosons with masses ranging from $10^{-24} \, \rm{eV} \apprle m \apprle 1 \, \rm{eV}$. The possible effects that ULDM may have on GW detectors can be looked for in how ULDM interacts with the mirrors of the interferometers. In fact, each ULDM model foresees a peculiar motion of the mirrors \citep{uldm1,uldm2,uldm3}. As a consequence, laser interferometers can probe the presence of DM by measuring the motion of mirrors caused by ULDM. Recent searches did not find any evidence of scalar ULDM in LIGO-Virgo data, still set upper limits on the coupling constant of scalar ULDM in the $[10^{-14},\, 10^{-11}] \, \mathrm{eV}$ range for the mass of the scalar field \citep{uldm1,uldm2,uldm3}.

\subsubsection{Dark Energy}
It is well established that 70\% of the total energy content of the Universe is composed of a mysterious form of energy, referred to as Dark Energy (DE), thought to be responsible for the accelerated expansion of the Universe.
Unlike normal matter or DM, DE does not clump or interact through gravity in the usual way. Instead, it seems to have a repulsive effect, pushing galaxies apart. Its nature is still unknown, but there are two main possibilities:
\begin{itemize}
    \item Cosmological constant: a constant energy density filling space uniformly, as introduced by Einstein in his equations of GR;
    \item Dynamic field (quintessence): a form of energy that can change over time and space.
\end{itemize}
Studying DE is one of the biggest challenges in modern cosmology, as understanding it could reveal fundamental insights into the fate of the Universe and physics beyond the Standard Model. GW detections can be used to probe the nature and existence of DE as well. In fact, as GWs can offer an independent method to determine distances without relying on traditional EM observations, they can provide precise constraints on cosmological parameters, including the Hubble constant and the EoS of DE itself.

\noindent The sensitivity of future GW detectors (Section \ref{future}) will also extend to high-redshift events. This feature will enable the study of the Universe's expansion history over a significant fraction of its age. By analyzing the distribution of GW sources across redshift, detectors can provide insights into the dynamics of DE and its influence on cosmic acceleration \citep{ET}.
Moreover, future GW detectors have the potential to observe the stochastic GW background, which may carry indirect information about DE and its role in the early Universe \citep{ET}.

\section{Complementarity with other experiments in the field}\label{compl}
Different GW detectors have been introduced over the years, and new experiments are being planned for the future to cover the GW spectrum as widely as possible. In fact, each GW detector has a limited sensitivity band according to its operating principle (how it is built, how it takes data, etc.), inside which it can observe GWs; therefore, by combining observations made with different GW experiments, one can broaden the range of observable frequencies with GWs and consequently explore new sources and physical phenomena.
These instruments can also operate in synergy with EM, neutrino, and cosmic ray detectors: each different emission provides different yet complementary information about astrophysical sources, providing a broad and unique way to probe the cosmos.
\subsection{Complementarity between GW detectors} \label{compl:GWs}
As we have mentioned in the previous Sections, the frequency of a GW is strictly linked to the process beneath its emission and to the source properties. By exploring different frequency ranges with GWs, we will therefore obtain different information and probe new systems and astrophysical sources. The frequency ranges we can detect GWs over are defined by the sensitivity range of the detector and thus by its operating principle. 

\begin{figure}[t]
	\centering
	\includegraphics[width=0.7\textwidth]{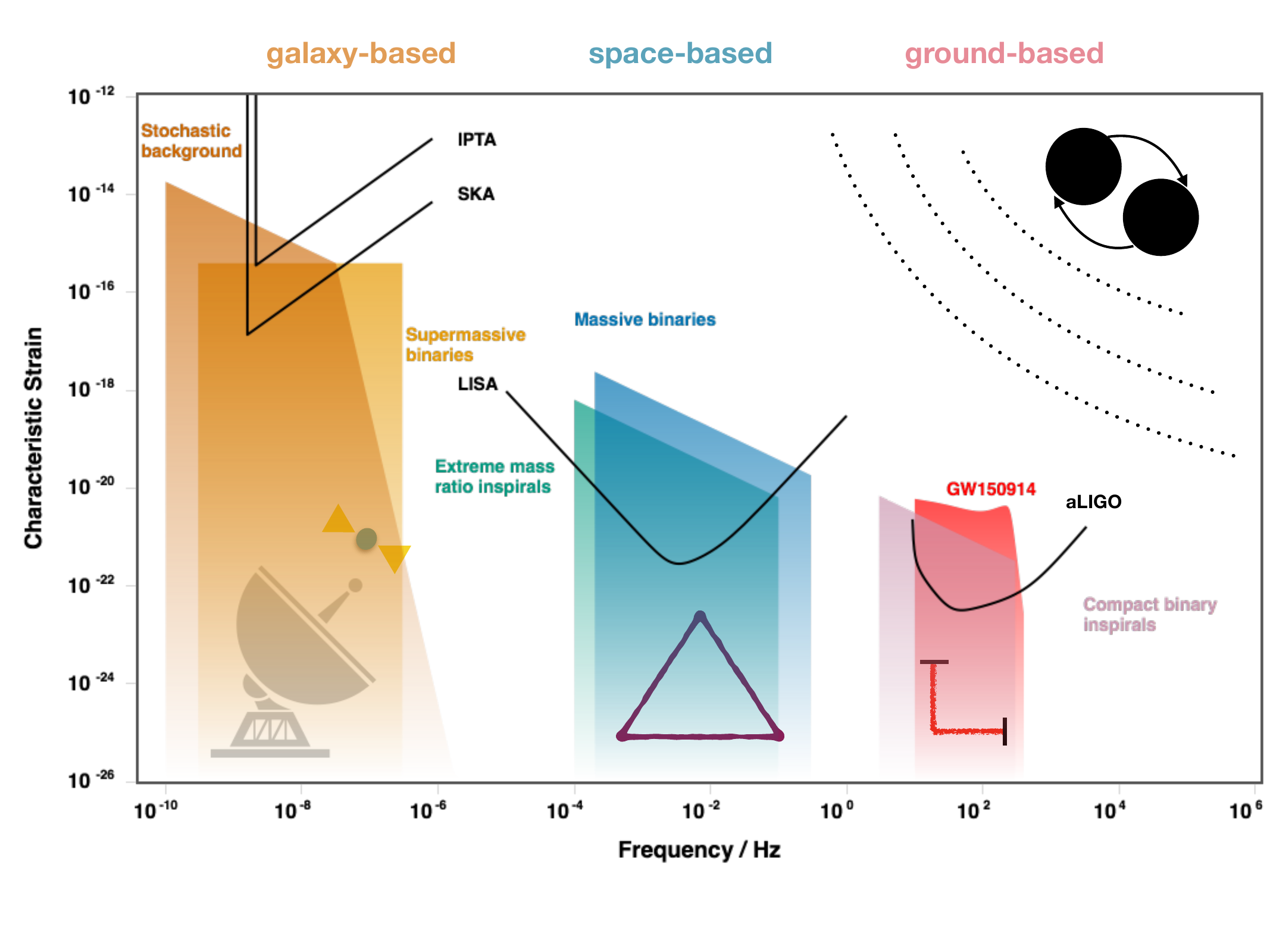}
	\caption{Expected amplitude strain through the GW spectrum.
    The shaded areas highlight the regions covered by different sources of GWs in the very-low, low, and high-frequency bands (see text). 
    The solid black lines are the sensitivity curves for some present and future experiments (Section \ref{future}). When the curves go below the shaded areas, the detectors are expected to be sensitive to that specific source. Credits: \citep{GWSpectrum}.
    }
    \label{fig:GWSpectrum}
\end{figure}

\noindent The GW spectrum can be roughly divided into three frequency ranges, each of which is expected to include GWs emitted from different systems (Figure \ref{fig:GWSpectrum}). As a consequence, different techniques and instruments are needed to cover the entire emission spectrum.
First, we have the high-frequency regime $\sim (1-10^4) \, \mathrm{Hz}$. GWs emitted at these frequencies can be emitted from CBCs of stellar BHs and/or NSs, asymmetric SN explosions, pulsar glitches, and asymmetrically rotating pulsars. These sources can be potentially detected with terrestrial interferometers, whose working principle has already been described in Section \ref{generic}.
The low frequency band is found at $\sim (10^{-5}-1) \, \mathrm{Hz}$. To make detections in this band, space-based interferometers are needed, which we will describe in Section \ref{compl:GWs:space}. Similar to the ground-based ones, these interferometers aim to detect GWs from the spatial displacement of test masses, but their location in space and construction properties allow them to reach higher sensitivities at lower frequencies. The sources expected in this range are mainly the coalescence of massive binaries, the inspiral of extreme-mass-ratio inspirals (EMRI), and the continuous emission from CBC systems far from coalescence that merge within the range of terrestrial interferometers.
Finally, we have the very low frequency band (VLF), referred to as the $\sim (10^{-10}-10^{-5}) \, \mathrm{Hz}$ range, approximately. This frequency range is covered by radio telescopes located on Earth, monitoring the arrival times of radio emissions from pulsars. In fact, a VLF GW is also a very-high-wavelength GW; therefore, to detect it, we need an apparatus that covers distances comparable to the wavelength of the GW. This is done by exploiting the typical distances $\sim 10^{17} \, \mathrm{km}$ between the Earth and galactic pulsars. In this frequency range, it would be possible to detect both GWs of cosmological origin (see Section \ref{physics:cosmology:universe}) and those produced from the inspiral of SMBH binaries.

\noindent The working principle of VLF GW detection is quite simple in practice. Since pulsars are highly magnetized and rapidly rotating NSs, their beamed EM emission can be detected on Earth as a succession of very stable radio pulses if the beam is not aligned with the rotation axis. When VLF GWs cross the spacetime between the Earth and the pulsar, the induced spacetime perturbations will modify the arrival times of the pulses, resulting in non-periodic contributions to their detections. Therefore, by tracking the arrival times of the EM pulses over very long periods of time $\Delta T$ and from a set of pulsars, we can observe and model any deviation from periodicity induced by GW frequencies up to $\sim 1/\Delta T$. This means that, to reach the very-low-frequency range, we require observing times of up to tens of years.

\noindent These detection techniques are also called Pulsar Timing Arrays (PTAs), or galaxy-based detectors, since they can span the size of the whole Galaxy. Moreover, several phenomena influence these time measurements, such as any orbital motion of the pulsar, the Earth's motion, general relativistic effects induced by massive objects in the traveling path of the photons, etc. All these effects need to be modeled a priori and taken into account in our measurements. In 2023, the first detection of very-low-frequency GWs was jointly made by four PTA collaborations around the world \citep{EPTA+InPTA, NANOGrav, PPTA, CPTA}. The amplitude of the detected signal was about $\sim 3 \times 10^{-15}$ at a frequency $f = 3\times10^{-8} \, \mathrm{Hz}$, consistent with the emission of GWs from a population of inspiraling SMBHs.

\begin{figure}[t]
	\centering
	\includegraphics[width=0.5\textwidth]{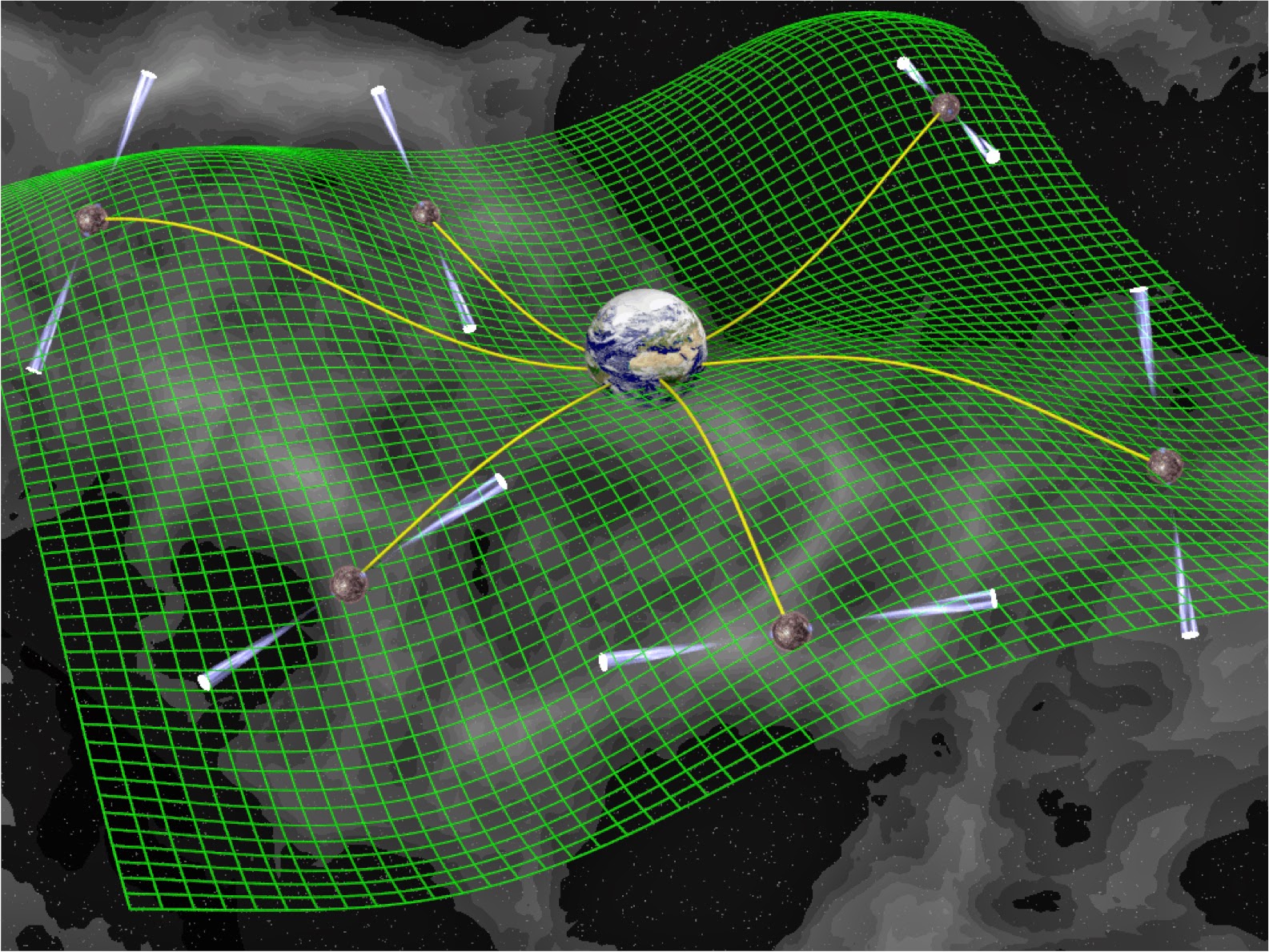}
	\caption{Artistic illustration of how the arrival times of pulsar radio flashes are affected by perturbed spacetime.
    Credits: David J. Champion}
    \label{fig:PTA}
\end{figure}

\subsection{GW170817 and the dawn of multimessenger astronomy} \label{compl:mma:170817}
In 2016 \citep{Branchesi_2016} outlined the feasibility and the potential of joint observations between GW detectors and other observatories, such as optical, gamma-ray, X-ray, cosmic ray, and neutrino observatories. Notably, certain categories of GW sources, transient sources in particular, have been observed or are expected to emit photons, neutrinos, and/or cosmic rays as well. The branch of astrophysics that deals with the detection of these multiple emissions, or messengers, from the same source is called multimessenger astrophysics, and it aims to reach an unprecedented level of understanding of astrophysical sources by combining the various information borne by each one of them. The idea is that each messenger is produced and emitted via different mechanisms and will therefore carry complementary information about the source that produced them. For instance, while the detection of EM radiation can help us identify the host galaxy of a source and therefore give us its precise localization, GW detections can inform us on the system's macroscopic properties, such as its mass, spin, spatial orientation, or its distance from us; neutrinos hold crucial information on the source's interiors and the nuclear processes that it undergoes. Finally, the emission of cosmic rays is strongly influenced by the acceleration processes and the environment (EM fields, composition, etc.) of the source. In this scenario, CBCs are thought to be, and have proved to be, very promising multimessenger sources: when the coalescence harbors at least one NS, EM emissions are expected, specifically, in the form of short Gamma-Ray Bursts (GRBs). GRBs are intense and highly variable beamed flashes of gamma-rays, followed by longer-lasting multiwavelength emissions in X-rays, optical, radio, and gamma-rays too. GRBs can be classified according to the duration of the initial gamma-ray emission, in short GRBs if it is $\apprle 2 \, \mathrm{s}$, and long GRBs otherwise \citep{GRB}.

\noindent The first joint GW-GRB detection happened on August 17, 2017, when the global network of the LIGO and Virgo interferometers observed for the very first time the merger of a BNS: GW170817 \citep{GW170817}. The event was later ($\sim 1.7 \, \mathrm{s}$) followed by the detection of the closest and least luminous GRB ever detected (GRB 170817A, \citep{GRB170817AFermi, GRB170817AINTEGRAL}). Thanks to the contribution of Virgo in the detection of the GW transient, a very precise sky localization for the event was obtained \citep{Gw170817properties}, which turned out to be crucial for the successful detections of the following multiwavelength EM emissions from the GRB: EM observatories were, in fact, able to be pointed in the identified region of the sky and successfully detect a bright optical transient as well as infrared, X-ray, and radio emissions that lasted several days after the BNS merger \citep{multim}. GW170817, therefore, marked the birth of multimessenger astrophysics with GWs and had profound implications for several aspects of physics, ranging from astrophysics to fundamental physics and even cosmology.

\paragraph{Astrophysics}
Before this first joint GW-GRB detection, theoretical works based solely on EM observations and population studies suggested a relation between the event powering the GRB emission and its duration. More specifically, CBCs are thought to be at the root of short GRBs, while the collapse of massive stars is expected for long GRBs \citep{GRBclassification}. In this scenario, the fact that GRB 170817A was classified as a short GRB and that it was unambiguously associated with a BNS merger confirmed how short GRBs can be produced by CBCs, as long predicted. 
Moreover, a theoretical model, known as the fireball model, had been proposed to explain the unique features of such emissions (rate of occurrence, energies emitted, spatial distribution, etc.) in relation to the physical processes underlying GRB emissions \citep{fireball}. By exploiting the duration of the GRB and the time delay between the two detections, it was possible to shed even more light and confirm the very details of the fireball model to an unprecedented level of precision \citep{GRB170817Aimplications}.\\
Probably, the most significant result related to the multimessenger observations of GW170817 comes from the optical, IR, and UV transients detected a few hours after the GW detection. The spectrum of these emissions was found to be consistent with a kilonova emission, i.e., a thermal optical/NIR transient emission that is thought to be produced by the radioactive decay of atomic nuclei, like gold and platinum, synthesized and then isotropically ejected during a CBC. This proved that BNS mergers are sites of neutron capture for the nucleosynthesis of heavy elements \citep{r-processes}. Finally, several days later, even X-ray and radio counterparts were detected, and, by monitoring the evolution of the light curves, it was also possible to constrain the size of the GRB beam and its displacement \citep{jet}.

\paragraph{Fundamental physics}
The $\sim 1.7 \, \mathrm{s}$ time delay between the GW and GRB detection was also used to put the first constraints on the propagation speed $v_{GW}$ of GWs to the speed of light. In fact, as GR predicts that GWs propagate at the speed of light, the joint detection provided an independent test of GR to those described in Section \ref{physics:GR}, which confirmed the predictions up to a level of $v_{GW}/c \, \sim 10^{-15}$ \citep{vGW}.

\paragraph{Cosmology}
Finally, the multimessenger detection was used to put cosmological constraints on the Hubble constant, as described in Section \ref{physics:cosmology:H}. In fact, while the optical, IR, and UV counterparts allowed for the identification of NCG 4993 as the host galaxy and a measurement of its redshift, the GW detection provided the distance of the CBC that powered the GRB emission.
By combining these values in the Hubble law \ref{eq:Hubble law}, a value of $H_0 = 70.0^{+12.0}_{-8.0} \, \mathrm{km} \, \mathrm{s}^{-1} \, \mathrm{Mpc}^{-1}$ was obtained \citep{H0170817}. This is one of the most relevant aspects of cosmological observations through GWs. In fact, we already mentioned how the two main measurements of the Hubble constant provide inconsistent values of $H_0$, originating the so-called Hubble tension, i.e., the discrepancy between a direct measurement of the Hubble constant and constraints from the early Universe. As a consequence, GW observations may help to shed new light on the origin of this discrepancy, pinpointing the source of the tension and eventually confirming the presence of new physics beyond the standard cosmological model.

\subsubsection{Other multimessenger sources}
Another example of a source that may benefit from multimessenger observations is the collapse of massive stars, which is associated with SN and long GRBs. For example, SN 1987A was observed through the detection of EM radiation \citep{SN1987Aem}, but also by two bursts of neutrinos consistent with the collapse of a massive star \citep{SN1987Aneutrino}. The detection of the neutrino counterpart was crucial for both confirming the explosion mechanisms modeled for Type II SN and in shedding more light on the nature and properties of neutrinos, such as their mass and flavour \citep{SN1987Amodels}.

\noindent Other relevant phenomena for multimessenger observations are NS instabilities, which are expected to be associated with the soft-gamma ray repeaters. These are anomalous X-ray pulsars that emit hard X-ray/gamma-ray, repetitive 0.1 sec flares, and occasionally giant flares. They are also associated with pulsar glitches, i.e., sudden increases in the NS rotational phase, frequency, or frequency derivatives observed in the radio and gamma-ray bands.

\noindent The EM sources mentioned above cover the whole EM spectrum, from high energies to radio, their emissions can be either beamed in jets or isotropic, and they can span different time scales from seconds to months, or even years. The broad emission band and the different timescales require a global network of ground-based telescopes and satellites to detect the EM signatures of the GW sources, and appropriate multi-messenger observational strategies as well as data analyses.

\section{Future development}\label{future}

As extensively discussed in the previous Sections, currently active GW detectors have validated some key astrophysical phenomena, such as the existence of compact objects and their detectability by ground-based detectors (Section \ref{physics:astro}). They also paved the way for multimessenger astronomy, as evidenced by GW170817 and the subsequent successful observational campaigns with partner observatories (Section \ref{compl:mma:170817}). However, the full scientific potential of GW observations remains untapped, especially in the low-frequency domain, for the post-merger dynamics, the detection of continuous GWs from isolated pulsars, GW emissions from SMBHs, and the stochastic GW background of both astrophysical and cosmological origin. For this reason, the GW community has been intensely working on both the next generation of ground-based detectors, which will improve performance with respect to the LVK network, and on unprecedented detectors based in space. In fact, as already mentioned, ground-based detectors cover frequencies approximately in the range $1 \, \mathrm{Hz} \apprle f \apprle 1 \, \mathrm{kHz}$, defined by the obvious size limitations of the instruments and by environmental noise; covering GW emissions in the $1 \, \mathrm{mHz} \apprle f \apprle 1 \, \mathrm{Hz}$ range, on the other hand, can only be achieved in space, with a constellation of satellites millions of kilometers apart.

\subsection{Ground based detectors} \label{future:ground}
\begin{figure}
\begin{minipage}{0.50\linewidth}
\centerline{\includegraphics[width=0.7\linewidth]{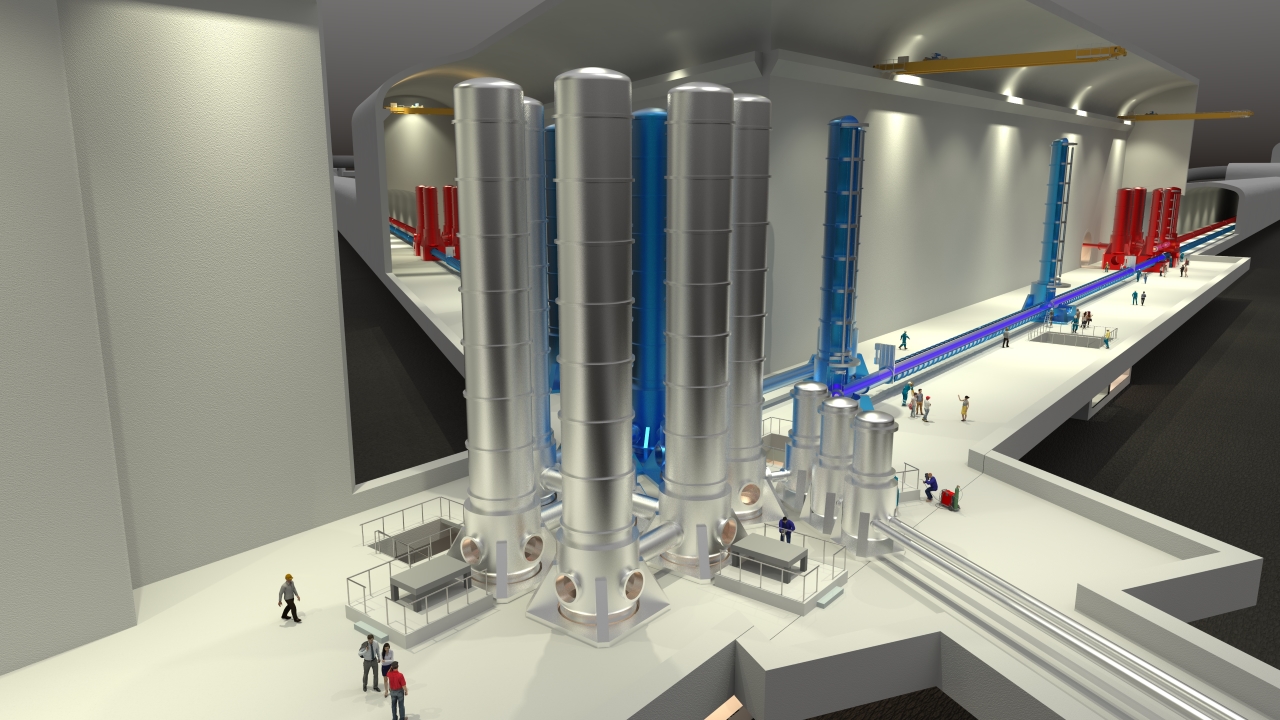}}
\end{minipage}
\hfill
\begin{minipage}{0.50\linewidth}
\centerline{\includegraphics[width=0.7\linewidth]{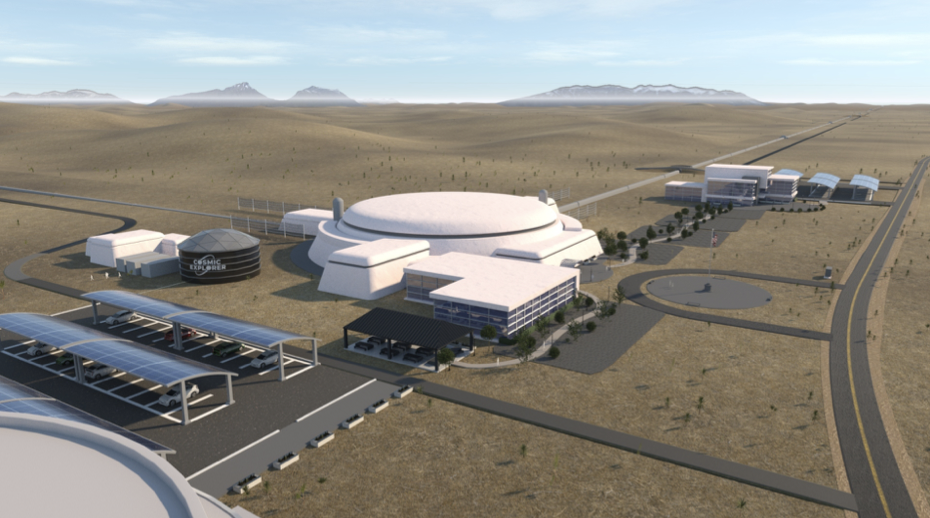}}
\end{minipage}
\caption[]{Artistic views of the underground infrastructure of ET \citep{ET2020} and CE \citep{cesite}.}
\label{fig:art}
\end{figure}

Parallel to the LVK successful observations, GW scientists have been developing a new generation of ground-based detectors as well as the construction of a third LIGO interferometer to be located in Maharashtra, India \citep{ligoindia}. These new-generation detectors (Figure \ref{fig:art}) consist of the Einstein Telescope (ET) \citep{ET2010, ET2011, ET2020} and Cosmic Explorer (CE) \citep{CE1, CE2, CE3, CE4}, which are promising to revolutionize the GW field by vastly enhancing the accessible bandwidth and sensitivity of ground-based detectors, extending our view of the Universe to unprecedented depths and details.

\subsubsection{Einstein Telescope} \label{future:ET}
\begin{figure}
\begin{minipage}{0.50\linewidth}
\centerline{\includegraphics[width=0.9\linewidth]{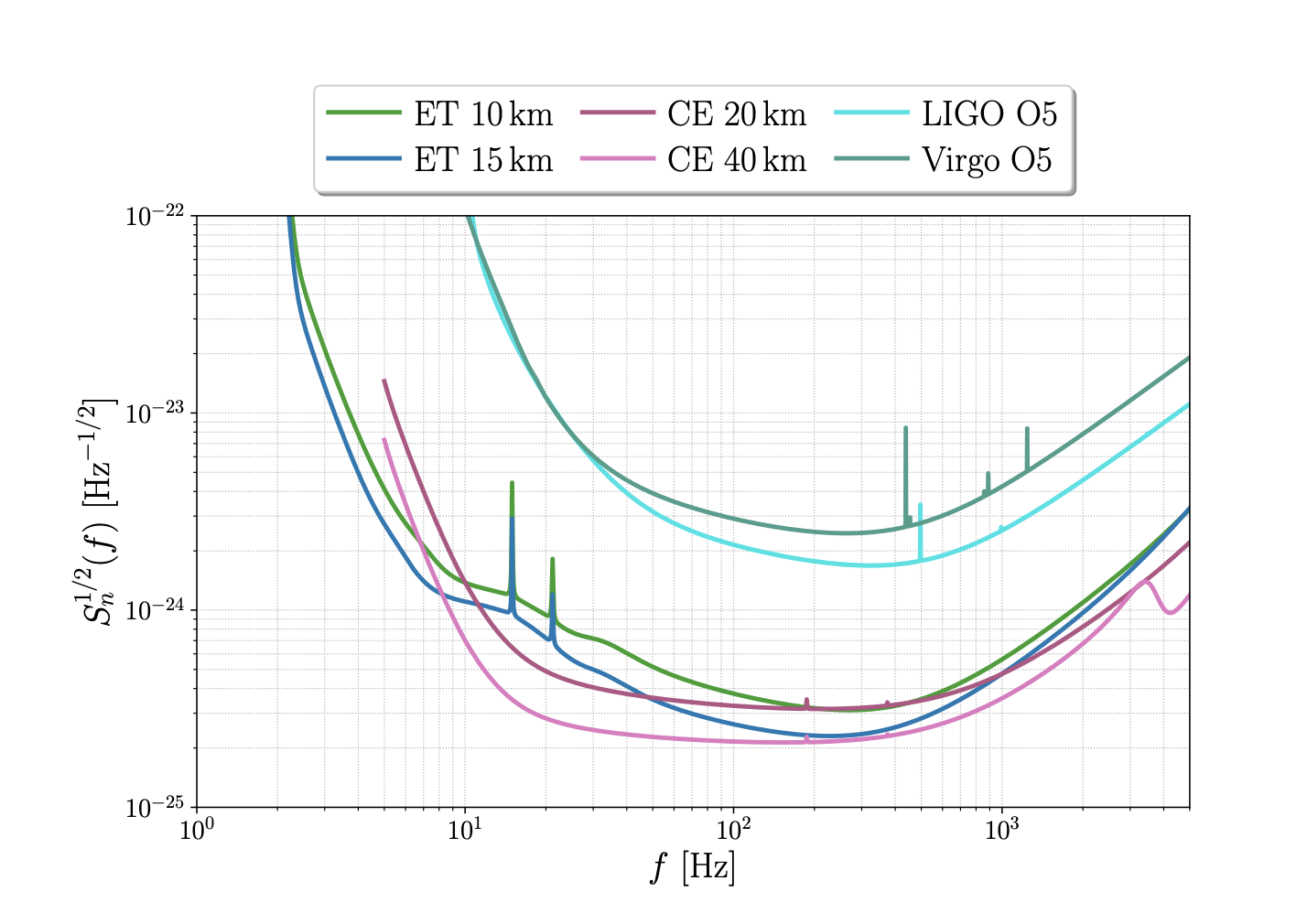}}
\end{minipage}
\hfill
\begin{minipage}{0.50\linewidth}
\centerline{\includegraphics[width=0.6\linewidth]{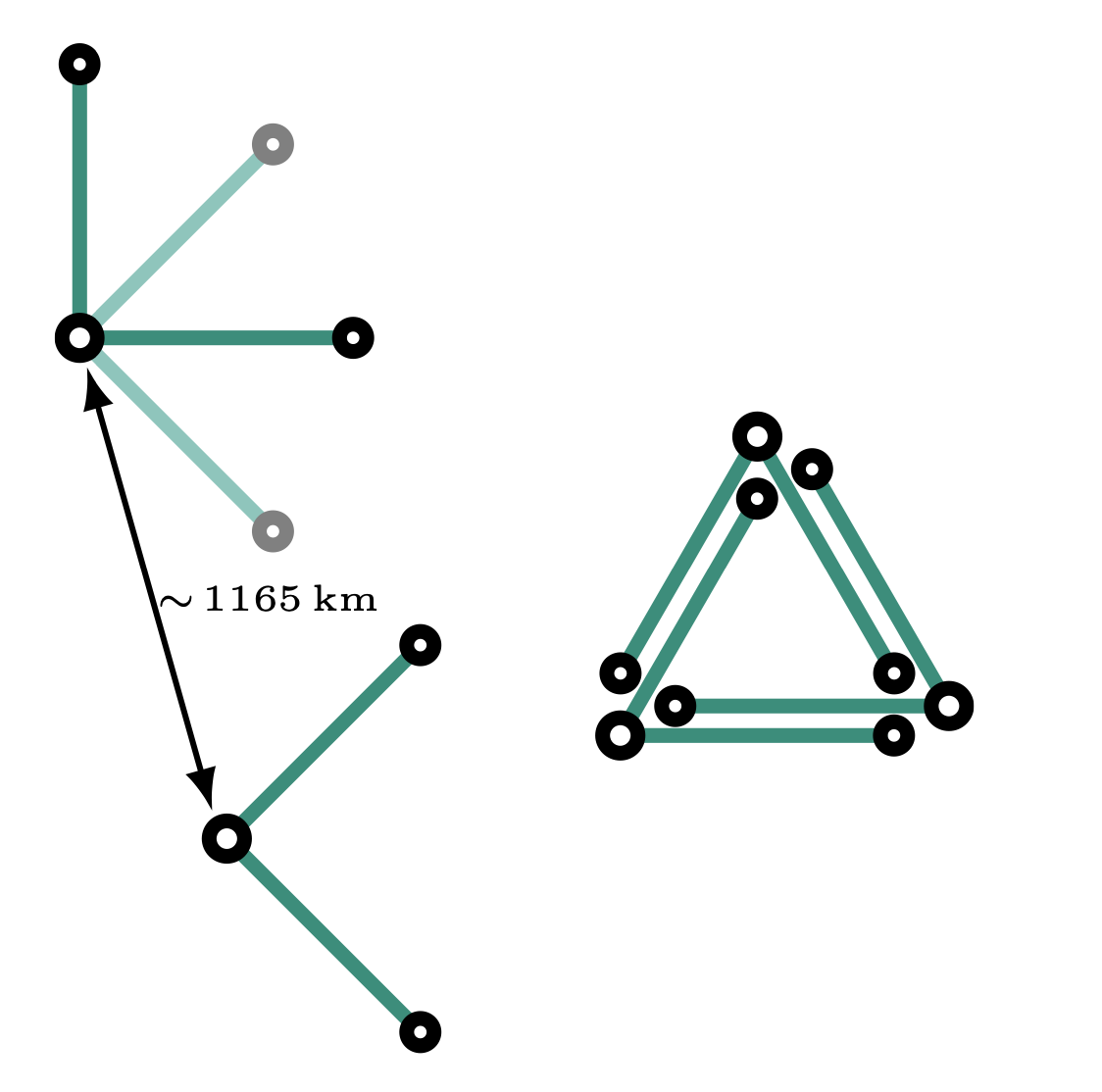}}
\end{minipage}
\caption[]{(left) ET and CE design sensitivity curves compared against design sensitivities of current and future GW detectors Credits: \citep{Iacovelli_2024}. (right) A schematic picture of the different geometries considered for ET: two widely separated L-shaped detectors (either parallel or at $45^{\rm o}$), or a triangle made of three nested detectors. Credits: \citep{coba}.}
\label{fig:comp}
\end{figure}

The Einstein Telescope is the European proposal for a next-generation GW detector, first proposed in 2010 \citep{ET2010}. The project consists of a ground-based interferometer, whose geometric configuration is still being debated: the most recent proposal consists of a detector network composed of two pairs of widely separated L-shaped interferometers with $15 \, \mathrm{km}$-long arms \citep{coba, Iacovelli_2024}, whereas the original project foresees three pairs of nested interferometers arranged in an equilateral triangle (Figure \ref{fig:comp}, right) with $10 \, \mathrm{km}$-long arms \citep{et1, ET2010, ET2011, ET2020}. For each interferometer pair, one detector is optimized for low frequencies (LF, $2 \, \mathrm{Hz} \apprle f \apprle 40 \, \mathrm{Hz}$) and the other for high frequencies (HF, $f \apprge 40 \, \mathrm{Hz}$, also called xylophone configuration). References \cite{coba, Iacovelli_2024} found that the difference between the two configurations in terms of the reachable science goals is minimal. The actual location of ET is still a matter of debate as well, with three sites competing to host ET: Sardinia (Italy) \citep{naticchionietal2014, digiovannietal2021, digiovanni2023, digiovanni2025}, the Euregio Meuse-Rhine between the Netherlands and Belgium \citep{koley19, koley2022surface, Bader_2022, vanBeveren_2023}, and the Lausitz region (Germany).

\noindent With ET, the lower accessible frequencies will be extended from $20 \, \mathrm{Hz}$ for current detectors, to $2 \, \mathrm{Hz}$, and the sensitivity will increase by a factor 8 across the band covered by current detectors \citep{ET2020} (Figure \ref{fig:comp}, left). To reduce seismic motion at the input of the suspension system of the mirrors and to reduce the impact of atmospheric disturbances \citep{hutt} and Newtonian noise \citep{harms,harms2022}, ET is also foreseen to be built underground at a currently planned depth between $250 \, \mathrm{m}$ and $300 \, \mathrm{m}$, regardless of the chosen configuration. Generally speaking, the extension of the bandwidth and the sharp increase in sensitivity will significantly improve the rate of detected events, giving the possibility to issue early warnings for CBCs several minutes before the merger, or even hours, depending on the source \citep{Branchesi_2016,Maggiore_2020,Nitz_2021,coba, Hu_2023}. In fact, the CBC detection rate for ET is expected to exceed $\sim 10^5$ events per year \citep{ET}. Moreover, CBC signals will spend more time in the accessible bandwidth of future GW ground-based detectors, thereby enabling early detections and accurate sky localizations, which favor multimessenger detections as well. As already mentioned in previous sections, multimessenger observations will be of primary importance in unveiling the secrets of some key astrophysical phenomena that are linked to the emission of GWs. The use of BNS mergers as bright sirens will also provide a more accurate estimation of the Hubble constant through GWs and will contribute to solving the Hubble tension enigma.

\noindent Overall, the performance of ET will offer unparalleled opportunities to probe the formation and evolution of stellar-mass BHs and NSs. By detecting events up to redshifts beyond $z = 10$, ET will allow us to map the population of compact objects, improving our understanding of stellar evolution, SN mechanisms, and the environments where compact binaries form \citep{ET}.

\noindent As far as binaries are concerned, ET will offer precise measurements of masses, spins, and tidal effects, offering insights into the EoS of NSs and the dynamics of BBH mergers. It will also investigate post-merger remnants and potential signals from long-lived NSs, as well as transient phenomena associated with glitches and instabilities. These observations, combined with EM and neutrino observations, will deliver a complete picture of these extreme astrophysical processes \citep{ET}.

\noindent The vast catalog of the detected GW events will enable more precise tests of GR in the strong-field regime as well. ET will probe deviations in inspiral, merger, and ringdown phases, search for exotic compact objects, and explore possible interactions between DM and GWs. These observations offer unique opportunities to detect signatures of physics beyond the Standard Model \citep{ET}.

\subsubsection{Cosmic Explorer} \label{future:CE}
The Cosmic Explorer was proposed by American institutions a few years after ET, and it foresees the construction of a network of two widely separated L-shaped interferometers, one with $40 \, \mathrm{km}$-long arms and the other with $20 \, \mathrm{km}$-long arms \citep{CE2,CE3,nsfreport}. There are also alternative scenarios for a single $40 \, \mathrm{km}$ facility, two $40 \, \mathrm{km}$ facilities, or even two $20 \, \mathrm{km}$ facilities \citep{CE3}. The baseline configuration will be widely based on the LIGO technology, including a $1064 \, \mathrm{nm}$ wavelength laser and, contrary to ET, room temperature operations. Other options are also under consideration for future upgrades. Since it is planned to be built on the surface, CE is also expected to adopt appropriate noise suppression and isolation systems to reach the design sensitivity goals at low frequencies.

\noindent The call for a $20 \, \mathrm{km}$-long facility, in addition to a $40 \, \mathrm{km}$-long facility, was motivated by the necessity of measuring the post-merger oscillations of BNS mergers which are expected to happen between $\sim (2-4) \, \mathrm{kHz}$, where the sensitivity of a $40 \, \mathrm{km}$-long detector increases with the travel time of the light down the arms \citep{CE3}. In particular, the $20 \, \mathrm{km}$ configuration could make use of “tuned” operation, i.e., the quantum noise is reduced in the $(2-4) \, \mathrm{kHz}$ band at the expense of higher quantum noise elsewhere. It was also found that a $20 \, \mathrm{km}$ detector yielded a $30\%$ improvement in the $(2-4) \, \mathrm{kHz}$ band sensitivity compared to a $40 \, \mathrm{km}$ detector, which would still address most science themes for its superior noise performance outside the same frequency range \citep{CE3}.

\subsubsection{ET and CE global network}

Although it does not match the current ET design sensitivity at VLFs (Figure \ref{fig:comp}), CE complements ET in terms of global coverage and redundancy. The collaborative operation of ET and CE as a global third-generation network will substantially improve event localization, potentially leading to uncertainties of $\sim 1 \, \deg^2$ for BBHs and $\sim 10 \, \deg^2$ for BNSs \citep{Iacovelli_2024,ET}, critical for triggering EM follow-ups in multimessenger astronomy. CE will also significantly contribute to key science goals, such as constraining merger rates, probing formation channels of SMBHs and IMBHs, and detecting potential PBH populations.


\subsection{GW detection in Space} \label{compl:GWs:space}

Ground-based interferometers, although capable of uncovering the secrets of GW emission from a few Hz to a few kHz, do not have access to sources that emit GW at frequencies below $\sim 1 \rm Hz$ due to the intrinsic limitations of terrestrial detectors. Nevertheless, below 1 Hz, many other GW sources that can help in the understanding of the Universe are expected: the continuous GW emission from BBH and BNS binaries far from coalescence, the hypothetical mergers of SMBH binaries, the plunge of a compact object into a SMBH, and merging white dwarfs in our Galaxy. Moreover, space-based detectors can be used in a complementary way to terrestrial interferometers. For example, the detection of the continuous emission from CBC systems far from coalescence can easily provide an estimate of their merger time and issue an accurate sky-localization. As a consequence, space-based interferometers can send alerts days, months, or even years in advance, in such a way that ground-based detectors and other partner observatories can prepare for the merger and increase the significance of their detection. For these reasons, the GW community began to think beyond what can be done with detectors located on the Earth's surface and started designing large-scale detectors that can operate in space.
\noindent Currently, two categories of detectors are designed to operate in space: 

\begin{itemize}
    \item an interferometer composed of a constellation of three identical spacecraft forming a near-equilateral triangle in a heliocentric orbit, transferring laser beams over distances of the order of millions of kilometers;
    \item a detector based on the Moon aimed at measuring the vibrational modes of our Satellite induced by the passage of a GW.
\end{itemize}

\subsubsection{Space based interferometers}

The operating principle of space-based interferometers is basically the same as terrestrial interferometers (Section \ref{generic}), but two main differences allow us to reach low frequencies. First, their dimensions are significantly greater because of the greater wavelengths to detect $\sim (10^{9}-10^{5}) \, \mathrm{km}$; secondly, they are located in space. This is both a necessity of the huge lengths covered by the apparatuses, but it also makes these interferometers immune to seismic noise artifacts that peak exactly as the frequency decreases. The combination of these two characteristics makes space-based interferometers excellent candidates for detecting low-frequency GWs.
At the moment, three experiments are aiming at building such an instrument: the Laser Interferometer Space Antenna (LISA), an international collaboration between the European Space Agency and NASA \citep{LISA,LISAreport}; the Deci-hertz Interferometer Gravitational wave Observatory (DECIGO), proposed by the Japanese Space Agency \citep{decigo}; and TianQin, a space-based interferometer being designed in China \citep{TianQin}.

\begin{figure}[t]
	\centering
	\includegraphics[width=0.3\textwidth]{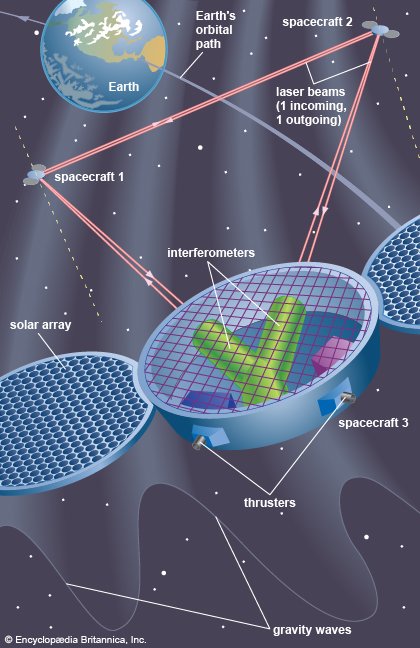}
	\caption{Artistic illustration of the LISA spacecraft.
    Image from: \href{https://www.britannica.com/science/gravitational-wave\#ref1070896}{Encyclopædia Britannica, Inc.}}
    \label{fig:LISA}
\end{figure}

\noindent Of the three, LISA is chronologically the first experiment ever proposed for the detection of GWs in space and is expected to be launched in 2037. In 2015, the LISA Pathfinder mission also successfully proved some of the operating principles of the detector \citep{LISAPathfinder}. LISA will cover frequencies in the range $(10^{-4}-1) \, \mathrm{Hz}$ and will be composed of three satellites separated by $2.5 \times 10^6 \, \mathrm{km}$, arranged to form an equilateral triangle. Each satellite will harbor two laser sources, which will be pointed toward the other two satellites, and two reflective mirrors, which will reflect the incoming beam (Figure \ref{fig:LISA}). This way, the spacecraft will work as a three-armed interferometer, detecting GWs from the minor displacements induced in the test masses. LISA is expected to detect GWs from various systems, including EMRI, massive binaries, and inspiraling CBCs, shedding more light on the formation channels of these binary systems and on the properties of massive objects \citep{LISAwf}. Moreover, thanks to its high sensitivity, we expect to cover higher distances and, therefore, use the GW detections to make cosmological constraints as well \citep{LISAreport, LISAcosmo}.

\subsubsection{Moon based detectors}

Earlier in this paper, we mentioned how the monitoring of vibrations in an elastic body excited by GWs was one of the first concepts proposed for detecting them \citep{Weber}. The detectors that rely on these operating principles are called resonant bar detectors (Section \ref{history}). Due to the dimensions of these bars (a few meters in length), the targeted signal frequencies were in the $\rm{kHz}$ range. When resonant bar detectors were first proposed, it became clear that monitoring the vibrations of a much larger body, such as the Earth or the Moon, could reveal GWs in the $\rm{mHz}$ band \citep{LGWA}. An attempt to prove this concept was the Lunar Surface Gravimeter experiment deployed on the Moon by the Apollo 17 crew \citep{LunarGravimeter,LunarGravimeter2}. Unfortunately, a technical failure greatly reduced the scientific scope of the experiment. Today, the same detector's concept is revived by the Lunar Gravitational-Wave Antenna (LGWA) \citep{LGWA,LGWAwhite}. LGWA, which is now in the design phase and is supported by the European Space Agency, foresees the deployment of a few sensors on the south pole of the Moon, which will be able to monitor the vibrations induced on the Moon by GWs in the $\rm{dHz}$ range. In fact, the LGWA frequency band will complement ground- and space-based detectors. It could therefore become an important partner observatory for joint observations with LISA, ET, and CE, while also contributing an independent science case \citep{LGWAwhite}.

\section{Conclusions}
\label{sec:conclusions}

This Chapter has been written a decade after the first direct detection of GWs, a breakthrough which was possible thanks to decades of experimental progress in interferometric detector technology. That discovery not only confirmed a key prediction of GR, but also inaugurated GW astronomy as a new observational science.
Over the past ten years, the global network of kilometer-scale interferometers, LIGO, Virgo, and KAGRA, has steadily advanced in sensitivity, enabling routine detections and a wealth of scientific results.

\noindent With hundreds of confirmed detections, the field has shifted from the study of single events to the statistical characterization of populations. This transition makes possible detailed studies of the formation and evolution of compact binaries, of their mass and spin distributions, and of their role in stellar evolution across cosmic time. These catalogs also provide empirical measurements of merger rates, offering a direct test for theoretical models of binary formation and dynamics.

\noindent GW observations have also opened a new window on the possibility of testing alternative theories of gravity. Unlike previous experiments, which were limited to weak gravitational fields, GW signals allow direct exploration of the strong-field regime near BHs and NSs. Analyses to date have confirmed the predictions of GR, while placing bounds on possible deviations predicted by alternative theories of gravity. As the detectors' sensitivities increase, these tests will become progressively more stringent, probing the nature of gravity under extreme conditions.

\noindent In cosmology, GW detections provide a new and independent method for measuring distances with respect to traditional EM observations. The joint GW-EM observation of GW170817 demonstrated the use of CBCs as standard sirens, yielding a direct measurement of the Hubble constant. As more detections are accumulated, both with and without EM counterparts, GW data will help address current discrepancies between early- and late-Universe measurements of the cosmic expansion rate. In the longer term, future detectors may observe a stochastic GW background of cosmological origin, revealing information about processes such as inflation or early phase transitions that are otherwise inaccessible.

\noindent The advent of multimessenger astronomy was another milestone. The coordinated follow-up of GW170817 confirmed the origin of short GRBs and provided a comprehensive and unprecedented view of the BNS merger process, including the detection of a kilonova. This highlights the value of combining different observational channels. As the detectors' duty cycles increase, similar joint campaigns will play a crucial role in characterizing a variety of transient sources.

\noindent Looking ahead, the prospects are even more compelling. Third-generation interferometers such as ET and CE, the forthcoming space mission LISA, and PTAs will extend the accessible frequency range and probe GWs across a broad spectrum of astrophysical and cosmological scales. These efforts will not only expand our astrophysical reach but also address fundamental physics questions ranging from the nature of gravity to the behavior of matter under extreme conditions and the nature of DM.

\noindent Altogether, these developments mark the beginning of what can be regarded as the golden era of GW science \citep{cerncourier}, in which gravitational observations will play a central role in shaping our understanding of the cosmos.

\bibliographystyle{harvard}
\bibliography{reference}

\end{document}